\documentclass[aps,prl,twocolumn,superscriptaddress,groupedaddress]{revtex4}  
\usepackage{graphicx}  
\usepackage{dcolumn}   
\usepackage{bm}        
\usepackage{amssymb}   
\usepackage{amsmath}
\usepackage{color}
 \usepackage{amsmath}
 \usepackage{amssymb}
\usepackage{graphicx}
\usepackage{bm}
\usepackage{color}
\usepackage{lmodern}

\usepackage{placeins}
\usepackage{inputenc}

\usepackage{tabularx}
\usepackage{hhline}
\usepackage{pifont}

\usepackage{graphicx}  
\usepackage{dcolumn}   
\usepackage{bm}        
\usepackage{amssymb}   

\begin{document}

\title{Deterministic remote entanglement of superconducting circuits through microwave two-photon transitions}
\author{P. Campagne-Ibarcq}
\email{philippe.campagne-ibarcq@yale.edu}
\affiliation{Department of Applied Physics, Yale University}
\author{E. Zalys-Geller, A. Narla, S. Shankar, P. Reinhold, L. Burkhart, C. Axline, W. Pfaff, L. Frunzio, R. J. Schoelkopf}
\affiliation{Department of Applied Physics, Yale University}
\author{M. H. Devoret}
\email{michel.devoret@yale.edu}
\affiliation{Department of Applied Physics, Yale University}
\date{\today}

\begin{abstract}
Large-scale quantum information processing networks will most probably require the entanglement of distant systems that do not interact directly. This can be done by performing entangling gates between standing information carriers, used as memories or local computational resources, and flying ones, acting as quantum buses. We report the deterministic entanglement of two remote transmon qubits by Raman stimulated emission and absorption of a traveling photon wavepacket. 
We achieve a Bell state fidelity of 73\%, well explained by losses in the transmission line and decoherence of each qubit.
\end{abstract}

\maketitle

\section{Introduction} Entanglement, which Schroedinger described as ``\emph{the} characteristic trait of quantum mechanics''~\cite{schroedinger}, is instrumental for quantum information science applications such as quantum cryptography and all the known pure-state quantum algorithms~\cite{jozsa2003role}. Two systems Alice and Bob that do not interact directly can be entangled if they interact locally with a third traveling system acting as a mediator. Since they can travel over long distances, photons are natural candidates for this role \cite{Hofmann2012}.\\
Remote entanglement was first demonstrated between two atomic clouds \cite{julsgaard2001experimental} traversed by a light beam measuring non-destructively a joint property.  This type of scheme was also used recently to entangle two superconducting circuits~\cite{Roch2014} but suffers from the difficulty to match both measurements to make the extracted information from one or the other system indistinguishable. Another scheme, widely used in atomic clouds \cite{matsukevich2006entanglement}, trapped ions \cite{Moehring2007},  solid-state spin qubits~\cite{Bernien2012}, quantum dots~\cite{Delteil2015} and superconducting circuits \cite{Narla2016} relies on the simultaneous emission of photons by both Alice and Bob, either through fluorescence or stimulated Raman emission. Entanglement is then heralded by detection of one of these photons, whose origin is erased by recombining them on a beam-splitter. This scheme is robust, in particular against photon losses, as long as the photons are made indistinguishable by matching the overlap of their wavepackets with the detector bandwidth. It should be possible  to entangle in this way two arbitrary nodes of a network for modular quantum computing~\cite{Kimble2008,Duan2010,Monroe2014}. 
But can we build an even simpler remote entangler, which would not require a which-path eraser and detector?\\
As depicted in Fig.~1a, a minimal protocol consists of entangling Alice with a propagating electromagnetic field whose state is then \emph{swapped} to Bob~\cite{Cirac1996}. For instance, we can stimulate an excitation of Alice that is conditional upon emission of a photon, which is then routed and absorbed by Bob. Such a scheme could also be used to shuffle information between the nodes of a network. However, the natural emission and absorption temporal envelopes of two identical nodes do not match as one is the time-reversed of the other. Following pioneering work in ion traps~\cite{keller2004continuous}, many experiments in circuit-QED have sought to modulate in time the effective coupling of the emitter to a transmission channel in order to shape the ``pitched'' wavepacket~\cite{Houck2007,Srinivasan2014,Pechal2014,Kindel2016,Pfaff2016}. Indeed, a rising exponential wavepacket could be efficiently absorbed ~\cite{Palomaki2012,Wenner2014,Pierre2014,Flurin2015} by the receiver. However, such a wavepacket in principle requires infinite dynamic range on the emitter coupling. A simpler approach consists of modulating both the emitter and receiver couplings in time to release and catch a time-symmetric wavepacket  \cite{Cirac1996, Korotkov2011}. While efforts were made in that direction~\cite{Yin2013, Andrews2015, Inomata2016}, deterministic entanglement between distant nodes has not been demonstrated with such a scheme so far~\footnote{While this experiment was performed,  Axline \emph{et. al.} (in preparation)  demonstrated efficient transfer of states and entanglement generation between two remote microwave resonators.}.\\
In this letter, we report entanglement generation between two distant transmon superconducting qubits by faithful release and capture of a time-symmetric wavepacket. Our scheme uses microwave pumps to  concurrently and coherently create a transmon excitation and a photon in a buffer resonator coupled to a transmission line~\cite{Kindel2016, Narla2016}. The photon is released, and after traveling along a $\sim1$~m cable and through microwave components, is captured by a second transmon qubit with a similar scheme. The entanglement purity is limited by photon losses in the line, which could be corrected for by purification~\cite{Bennett1995, Kalb2017}, and intrinsic decoherence of each qubit, which could also be improved.\\

\section{Driving a two-photon transition} 
\begin{figure}[]
\includegraphics[scale=0.6]{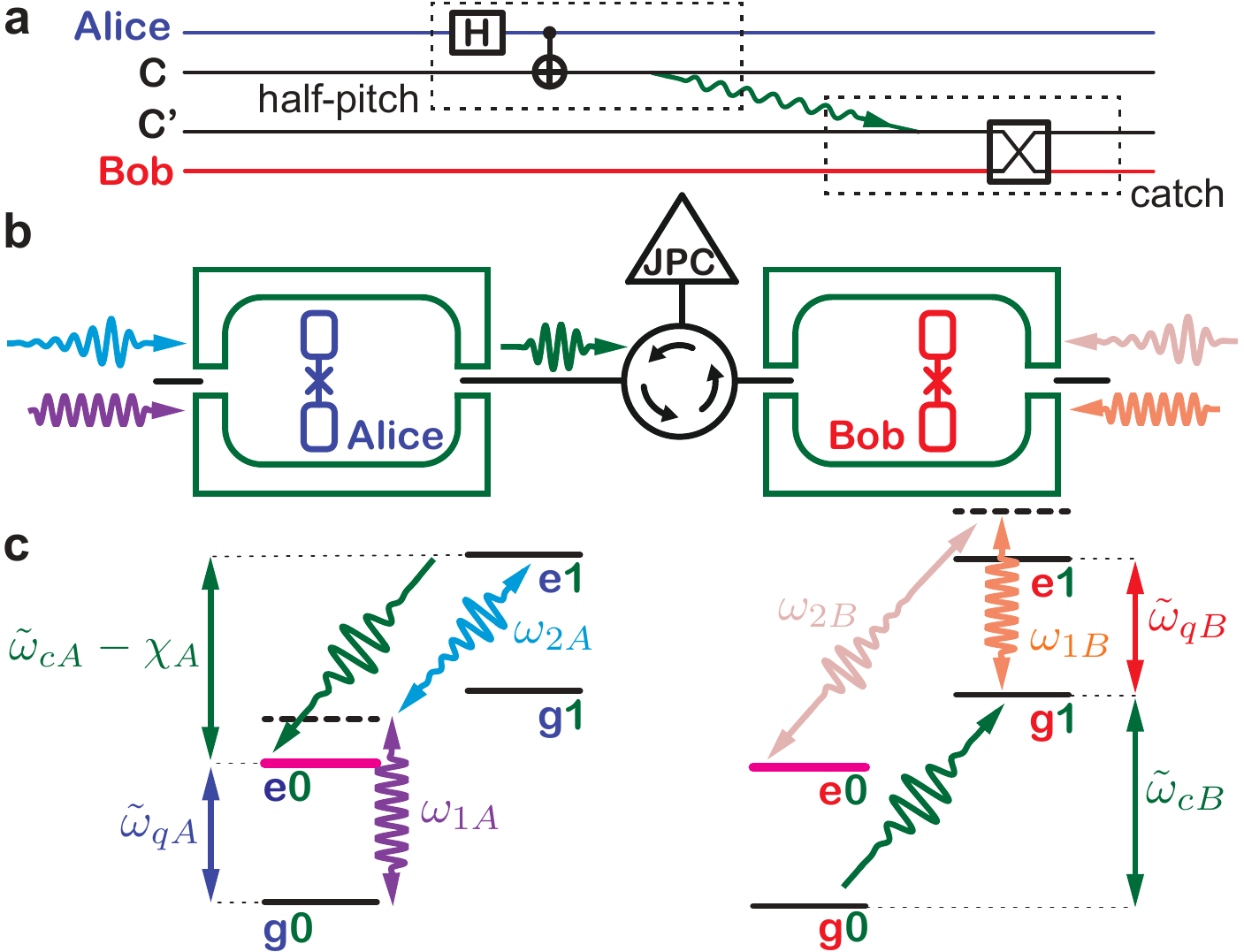}
	\caption{\label{fig:setup} \textbf{a)} Minimal logical circuit for remote entanglement. Alice is entangled with the ancillary system C by a Hadamard and a CNOT gate. The information propagates to C' (green wave) where it is \emph{swapped} to Bob.  \textbf{b)} Setup schematics and \textbf{c)} energy level diagram. Two transmon qubits Alice (in dark blue, dressed frequency $\tilde{\omega}_{qA}$, see text for details) and Bob (in red, dressed frequency $\tilde{\omega}_{qB}$) are dispersively coupled to two resonant cavities (in green, dispersive couplings $\chi_{A,B}$). The cavities lowest energy modes are frequency matched ($\tilde{\omega}_{cA}-\chi_A\simeq \tilde{\omega}_{cB}$) and are strongly coupled to a directional transmission line routing photons from Alice to Bob. By simultaneously driving Alice (Bob) with the detuned purple microwave at $\omega_{1A}$  (orange, at $\omega_{1B}$)  and her cavity with the detuned light blue microwave at $\omega_{2A}$  (light pink, at $\omega_{2B}$), we drive a Raman-type two-photon transition. For Alice, we choose $\omega_{1A}+\omega_{2A}=\tilde{\omega}_{qA}+\tilde{\omega}_{cA}-\chi_A$ to resonantly drive $|g0\rangle \leftrightarrow |e1\rangle$ (see (c) left diagram). A photon can  eventually be emitted in the line (green wave). The wavepacket is shaped by modulating the pump amplitude. This photon is absorbed by Bob by driving $|g1\rangle \leftrightarrow |e0\rangle$ with $\omega_{2B}-\omega_{1B}=\tilde{\omega}_{cB}-\tilde{\omega}_{qB}$ (right diagram). After a full photon pitch and catch, the system is in $|e0\rangle_A |e0\rangle_B$ (in magenta). After a ``half'' pitch, the qubits are entangled. 
}
\end{figure}

The experimental setup is schematically depicted in Fig.~\ref{fig:setup}b. Two superconducting transmon  qubits~\cite{Paik2011}, Alice and Bob, are embedded in two indium-plated copper cavities, anchored to the base stage of a dilution refrigerator (see~\cite{supmat,Narla2016} for device fabrication and setup details). The photon damping rate $\kappa=2\pi\times1~\mathrm{MHz}$ for the lowest energy mode of each cavity  is set by relaxation through a well-coupled port into a common microwave transmission line, which dominates over both the internal losses and relaxation through a second port. This last port is used to apply resonant microwave drives  to perform control operations on a single mode, such as qubit rotations  at $\omega_{qA,qB}/2\pi  \sim 5~\mathrm{GHz}$, or cavity displacements at $\omega_{cA,cB}/2\pi\sim 7.5~\mathrm{GHz}$. Interestingly, we can also directly drive common two-excitation transitions of these modes such as $|g0\rangle \leftrightarrow |e1\rangle$ or $|g1\rangle \leftrightarrow |e0\rangle$~\cite{wallraff2007sideband}. Here $|0\rangle$ and $|1\rangle$ designate Fock states of the cavity and $|g\rangle$ and $|e\rangle$ the ground and first excited states of the qubit.  This is done by  simultaneously applying a sideband pump at $\omega_{1A,1B}$ detuned from the qubit frequency by $\Delta/2\pi=100~\mathrm{MHz}$ (purple and orange waves on Fig~1) and another at $\omega_{2A,2B}$ detuned from the cavity frequency by $\pm\Delta$ (light blue and pink waves).\\


Let us consider separately each system Alice \emph{or} Bob. One can show~\cite{mirrahimi2014dynamically, leghtas2015confining, supmat} that in a displaced frame and using a rotating wave approximation, the system Hamiltonian in presence of pumps at $\omega_1$ and $\omega_2$ reads
\begin{equation}
\begin{split}
\frac{H}{\hbar}=&\tilde{\omega}_q(t) q^{\dagger}q+\tilde{\omega}_c(t) c^{\dagger}c -\frac{\alpha}{2}(q^{\dagger}q)^2 -\chi q^{\dagger}q c^{\dagger}c \\
&~~~+e^{-i (\omega_1+\omega_2)  t} g_{\mathrm{s}}(t)~q^{\dagger} c^{\dagger} + h.c. \\
&~~~+e^{-i (\omega_1-\omega_2) t} g_{\mathrm{c}}(t)~q^{\dagger} c + h.c.
\label{hamil}
\end{split}
\end{equation}
where $c$ and $q$ are the annihilation operators for the cavity and qubit modes, $\alpha$ is the anharmonicity of the transmon mode, $\chi$ the dispersive shift~\cite{blais2004cavity}, and  $\tilde{\omega}_q(t)$ and $\tilde{\omega}_c(t)$ are the Stark shifted frequencies of the transmon and cavity modes in presence of the pumps. These dressed frequencies and the \emph{squeezing}  and \emph{conversion}  strengths $g_\mathrm{s}(t)$ and $g_\mathrm{c}(t)$ are slow varying compared to $\Delta$ and read
\begin{subequations}
\begin{align}
\tilde{\omega}_q=&~\omega_q -\chi |\xi_2|^2 -2\alpha |\xi_1|^2 \label{starkq}\\
\tilde{\omega}_c=&~\omega_c  -\chi |\xi_1|^2\\ 
g_{\mathrm{s}}=&~\chi \xi_1 \xi_2 \label{squeeze}\\ 
g_{\mathrm{c}}=&~\chi \xi_1 \xi_2^{\ast} \label{conversion}
\end{align}
\label{xis}
\end{subequations}
Here, $\omega_q$ and $\omega_c$ are the frequencies in the absence of the pumps. $\xi_1$ and $\xi_2$ are the effective pump amplitudes -which correspond to the frame displacements used to get to Eq.~(1)- and are proportional to the amplitude of the pump tones. Note that since the cavity mode is only weakly anharmonic, we have neglected a frequency shift of the cavity mode proportional to $|\xi_2|^2$~\cite{supmat} .\\
The conversion or squeezing process (red or blue sideband) can be selected by setting either
\begin{subequations}\label{resonances}
\begin{align}
\tilde{\omega}_q +\tilde{\omega}_c-\chi&=\omega_1+\omega_2 \label{res1} & \rightarrow&~~~~ |g0\rangle \leftrightarrow |e1\rangle \\
\tilde{\omega}_q -\tilde{\omega}_c&=\omega_1-\omega_2 &\rightarrow&~~~~ |g1\rangle \leftrightarrow |e0\rangle \label{res2} 
\end{align}
\label{resonances}
\end{subequations}\label{resonances}
in driving the two-photon transition.
\begin{figure}[]
\includegraphics[scale=0.4]{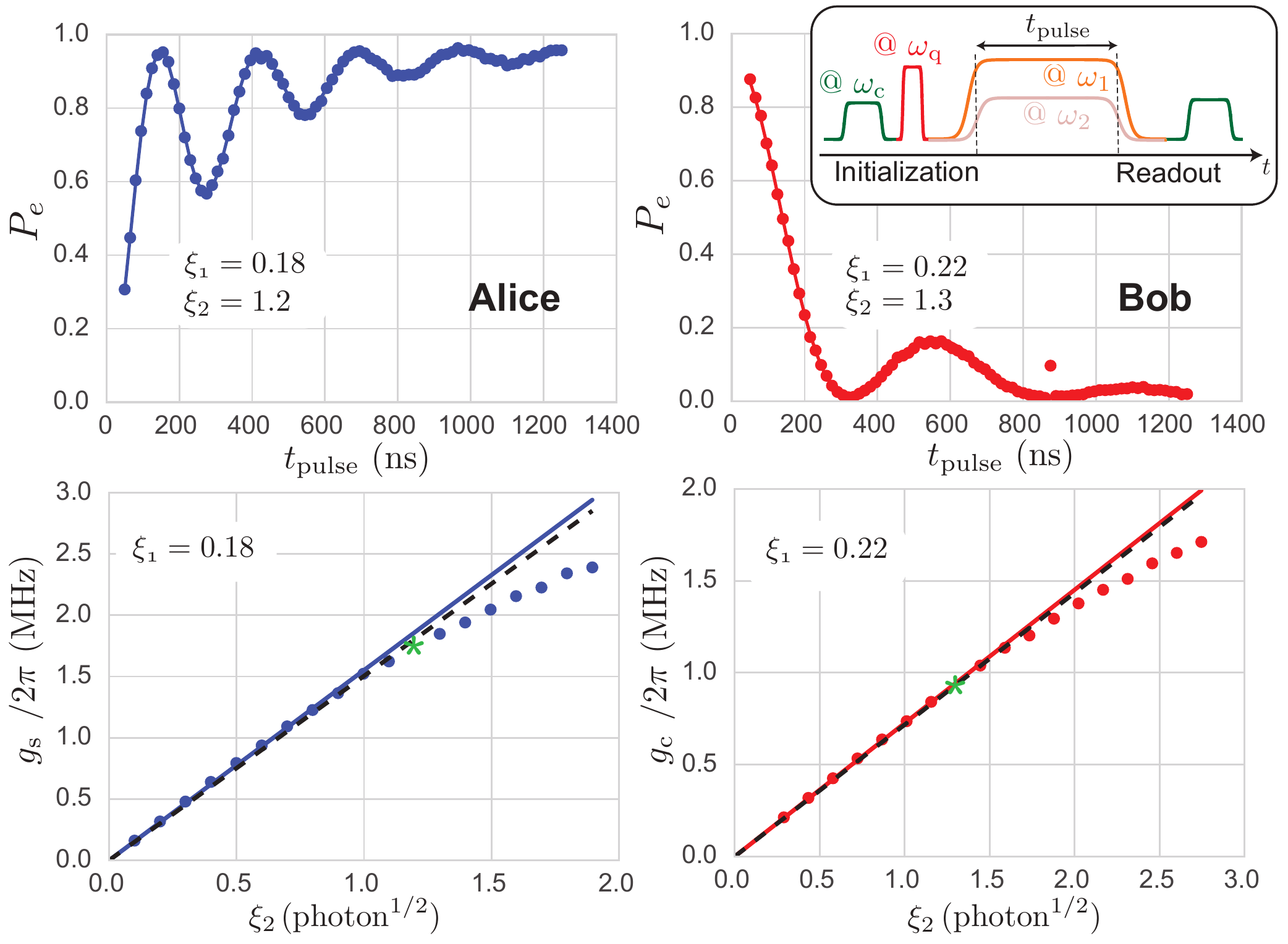}
\caption{\label{fig:rabi}   \textbf{Top panels} Rabi oscillations when driving a two-photon transition for a varying duration $t_{\mathrm{pulse}}$  are recorded in the qubit excited state populations (dots). Alice is initialized in $|g\rangle$ and Bob in $|e\rangle$. The pump amplitude values $\xi_1$ and $\xi_2$ are calibrated through Stark-shift measurements (see text and \cite{supmat}). Lines  are fits for the two-photon drive strengths $g_\mathrm{s}$ and  $g_\mathrm{c}$. \textbf{Inset: }Pulse sequence schematics. Pump pulse edges are smoothed to 128~ns and the pump~1 pulse is 100~ns longer. \textbf{Bottom panels} The extracted drive strengths are plotted when varying $\xi_2$ (dots, the green stars are from the top panel fits).  For each point, the cavity pump frequency is tuned to match the resonance condition Eq.~(3). Lines are linear fits of the non-saturated regions and their slopes are used as a calibration for the release and capture of a shaped photon. Dashed black lines are the drive strengths $|g_{\mathrm{c},\mathrm{s}}|=\chi |\xi_1 \xi_2|$ predicted from Stark-shift calibration of $\xi_{1,2}$~\cite{supmat}.
}
\end{figure}
The resonance condition Eq.~\eqref{res1} is used for Alice. As shown by the energy-level diagram of Fig.~1, this pumping, combined with the cavity dissipation, eventually brings the system to the state $|e0\rangle$ (highlighted in magenta). If the qubit is initially in $|g\rangle$, a photon is emitted in the line (green wave). Conversely, the resonance condition Eq.~\eqref{res2} is used for Bob, and if the qubit is initially in $|g\rangle$, it can absorb the incoming photon and excite to $|e\rangle$ (level highlighted in magenta), provided that the photon is resonant with the cavity frequency. This is made possible by designing the two cavities so that their transition nearly match ($(\omega_{cA}-\chi-\omega_{cB})/2\pi= 600~\mathrm{kHz}$), and compensating for the remaining mismatch with appropriately modulated pumps (see Fig.~3). Indeed, modulating the amplitude and frequency of the pumps in time makes it possible to choose the shape of the wavepacket containing the emitted photon, and to capture it efficiently. Accurate control of the drive strengths while matching the resonance conditions is the main difficulty of this experiment. \\ 
First, we must determine the unknown scaling factor linking the amplitude of the applied pumps to the effective amplitudes $\xi_{1A, 2A, 1B, 2B}$. This is done by measuring the shift of the qubit transition peaks in presence of the pumps and using Eq.~\eqref{starkq}, or any other quantity predicted by Eqs.~\eqref{xis}. Such spectroscopic measurements are presented in the supplementary materials~\cite{supmat}. While the Stark shifts display a characteristic linear dependence in the pump powers, some of the predictions from Eqs.~\eqref{xis} do not agree quantitatively.  In practice, we use an empirical approach. The amplitude of the two-photon drives being determined by the product of the pump amplitudes, we set $\xi_1$ and $\omega_1$ at a constant value. The cavity frequency is fixed (see Eq.~(2b)), and so is the frequency of the released photon. To vary $g_\mathrm{s}$ or $g_\mathrm{c}$, we then simply varies  $\xi_2$ and change accordingly the frequency $\omega_2$ to fulfill the resonance condition~(3). Note that when doing so, the dispersion of the transmission lines needs to be calibrated out  to keep accurate control of $\xi_2$, which can be done by using the qubit as an \emph{in situ} spectrometer~\cite{supmat}.\\
Following this protocol, we record Rabi oscillations of these two-photon transitions, presented on Fig.~2. The qubits are first initialized in $|g\rangle$ (Alice) or $|e\rangle$ (Bob)  by single-shot dispersive measurement  using a near quantum limited Josephson Parametric Converter~\cite{Bergeal2010b, Roch2012} (JPC) and fast feedback control~\cite{Riste2012, Campagne-Ibarcq2013}. The two-photon transition is then driven by  simultaneously applying the two pump tones for a varying time length $t_\mathrm{pulse}$. For Alice, we record an oscillation in the excited state population decaying to 1 at a rate $\kappa$, as $|e\rangle$ is a dark state in presence of cavity dissipation (see Fig.~1). Note that the edges of the pulses are smoothed as depicted in the top right inset so that the oscillation does not start at $P_e=0$. We can fit this oscillation by solving a quantum Langevin equation~\cite{gardiner2004quantum, supmat} on the qubit and cavity modes and the value of $g_\mathrm{s}$ as a single fit parameter. Inversely, for Bob (right panel), the excited state population decays to 0.  Note that this feature can be used for efficient  cooling of the qubits before the experiment~\cite{supmat}. In both cases, we then repeat the measurement when varying $\xi_2$. The extracted values of $g_\mathrm{s}$ and $g_\mathrm{c}$ display the expected linear dependence at low pump power (lines are linear fits) and are in good agreement with predictions from Eqs.~(\ref{squeeze},\ref{conversion}) with the values of $\xi_1$, $\xi_2$ and dispersive shifts $\chi_{A}/2\pi=8.36~\mathrm{MHz}$, $\chi_{B}/2\pi= 3.26~\mathrm{MHz}$ extracted from spectroscopic measurements~\cite{supmat} (dashed black lines). This behavior validates our model and gives an accurate calibration of the drive strengths. Saturation at large drive amplitudes is mainly due to saturation of the mixers used to generate the pulses. Note that our model also neglected some non-linear effects such as the anharmonicity inherited by the cavity mode~\cite{supmat} and the non confining nature of the transmon \emph{cosine} potential. For the actual release and capture presented in next sections, we use smaller values of $\xi_1 < 0.1$ and $\xi_2 < 0.25$ (see~\cite{supmat} for the corresponding Rabi oscillations) as the qubit coherence times were degraded at larger drive amplitude. This unexpected effect may originate from the aforementionned non idealities, compounded by the small pump detuning $\Delta$ -limited by our pulse generation scheme~(see Fig.~S1 in \cite{supmat})- compared to the transmon anharmonicity ($\Delta < \alpha_{A,B}\sim2\pi \times200~\mathrm{MHz}$). \\

\section{Excitation transfer} With the drive strengths calibrated, we now want to generate a photon with Alice and capture it with Bob. Following Cirac \emph{et al.}, we note that if we find some control sequence in time for $g_\mathrm{s}$ that, when applied to Alice, triggers the release of a photon in a wavepacket of given shape, the time-reversed sequence would succeed in capturing an incoming photon in a reversed wavepacket. Since Alice and Bob have similar properties (dissipation rates and maximum two-photon drive strengths), it is then natural to choose a time-symmetric shape for the traveling photon, and time-reversed controls for Alice and Bob. \\
\begin{figure}[]
\includegraphics[scale=0.37]{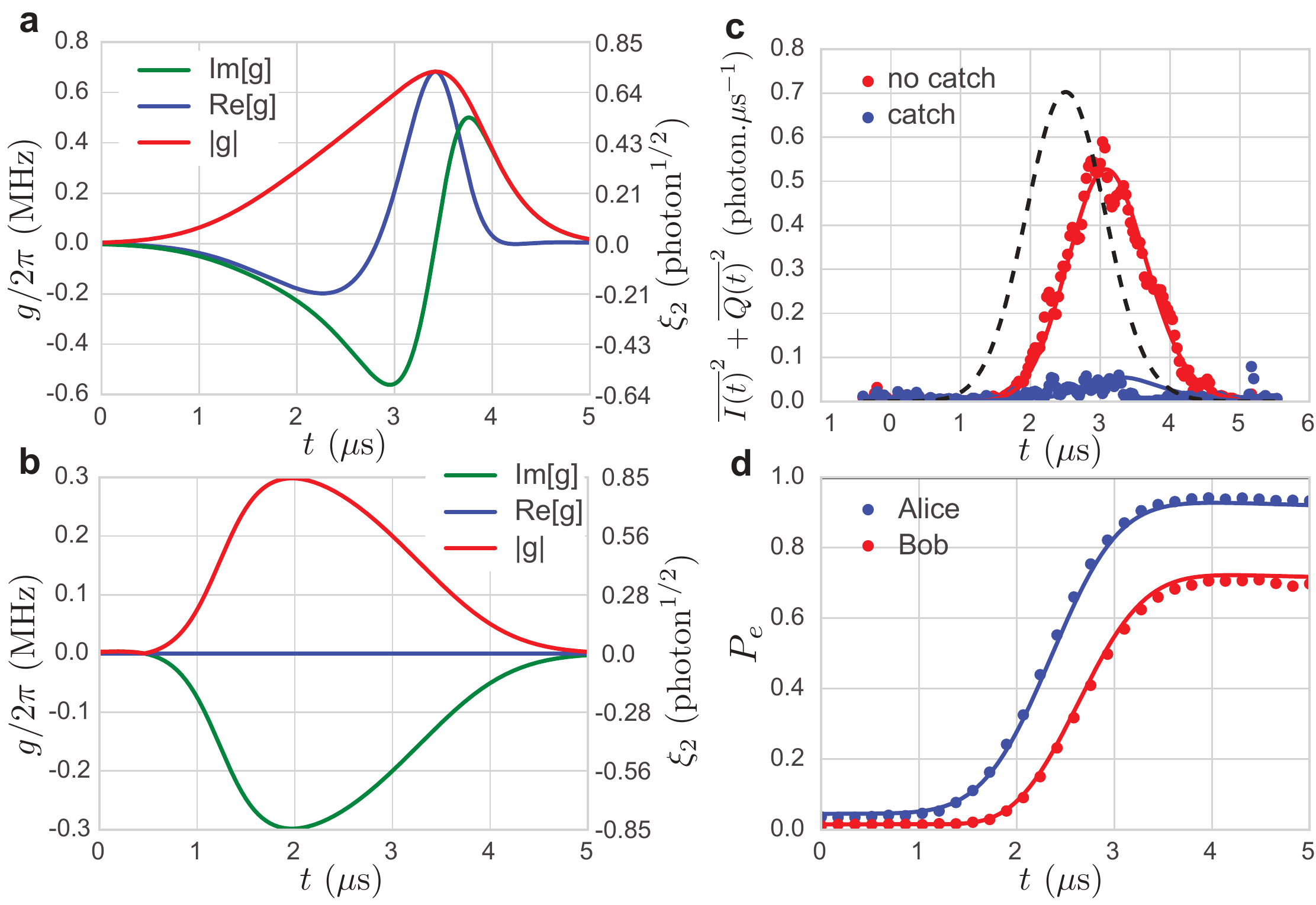}
\caption{\label{fig:full}  \textbf{Release and capture of a shaped photon. a)} Calculated complex amplitude of the drive strength for Alice to emit a photon in a Gaussian traveling mode with deviation $\sigma=800$~ns  and \textbf{b)} for Bob to fully capture this same photon. The traveling mode is chosen at Bob's cavity resonance frequency. These controls are realized by holding $\xi_1$ constant and varying $\xi_2$ as represented on the right axis. \textbf{c)}  When releasing the same wavepacket in the phase-defined state $(|0\rangle+|1\rangle)/\sqrt{2}$ (same sequence but Alice prepared in ($|g\rangle+|e\rangle)/\sqrt{2}$), we record the square value of the mean field reflected off from Bob's cavity  when capturing the traveling photon ($g_{\mathrm{c}}$ as in (b), blue dots) or not ($g_{\mathrm{c}}=0$, red dots). Lines are predictions from cascaded quantum system simulation including all imperfections and data is scaled by a unique value accounting for the uncalibrated amplification of the measurement chain. The reflected wavepacket is shrunk (due to finite line transmission) and delayed compared to the one predicted to travel from Alice to Bob (black dashed curve). \textbf{d)} Excited state populations of Alice and Bob during the transfer (dots), measured by interrupting the transfer control pulses after a duration $t$ and subsequent dispersive readout of the qubits. Lines are predictions from simulation with the same parameters as (c).
  }
\end{figure}
We choose a Gaussian shape for the traveling wavepacket. The controls required for releasing and capturing a photon in this traveling mode are computed using a method adapted from~\cite{Korotkov2011} and described in detail in~\cite{supmat}. The main idea is that if the photon capture is perfect, the reflected field from Bob is always in vacuum and the input-output relations linking the standing cavity modes to the right-traveling wave amplitude $a_{\mathrm{t}}$ in the transmission line read $c_{A}=a_{\mathrm{t}}/\sqrt{\kappa}=-c_{B}$ (we neglect propagation delays and absorption in the line). 
Thus, imposing the instantaneous photon flux $a^{\dagger}_{\mathrm{t}}a_{\mathrm{t}}$ to be Gaussian and to contain one photon sets the evolution of $c_{A}$ and $c_{B}$. Given the initial state of the qubits and the Hamiltonian~\eqref{hamil}, this determines the  choice of $g_\mathrm{s}$ and $g_\mathrm{c}$, represented on Fig~3a,b. Note that beyond these slowly varying envelopes, the pulses are modulated at $\omega_2$ and chirped to match the resonance conditions Eq.~(2) at all times.  The characteristic time $\sigma=800~\mathrm{ns}$ of the traveling wavepacket was chosen as short as possible to limit decoherence and relaxation of the qubits during the protocol while keeping  $g_\mathrm{s}$ and $g_\mathrm{c}$ below values where they display unpredicted non-linear dependance in the pump amplitudes (see Fig.~2). \\
It is important to note that Alice and Bob's control are not time-symmetric of one another. Indeed, to compensate for the small mismatch between Alice and Bob cavity lines (600~kHz when dressed by their respective qubit pumps), one needs to slightly modify one of the resonance conditions~Eq.~(3), which adds a small detuning in Alice's or Bob's Hamiltonian when expressed in a common rotating frame. We choose to modify Alice's resonance condition, so that the traveling photon is at Bob's cavity resonance frequency. The resulting control $g_s$ is slowly rotating and has a larger amplitude to compensate for this detuning. Thus, it is possible to choose to some extent the frequency of the released photon, and symmetrically, it would be possible to catch a wavepacket not resonant with Bob's cavity. However, pitching or catching a photon far out of the cavity bandwidth requires larger drive amplitudes. \\
The photon transfer is validated by measuring the qubit populations in time (Fig.~3d), which reveals a transfer efficiency of $70~\%$, when not correcting for any  experimental imperfections. These being independently calibrated~\cite{supmat}, one can reproduce the results with full cascaded quantum system simulations~\cite{gardiner2004quantum} (lines). The dominant error sources are finite readout fidelity (7~\% error),  decoherence of the qubits (11~\% error) and photon loss in the line (15~\% error). Since this last figure is poorly constrained by our calibration~\cite{supmat}, we also perform a direct heterodyne detection of the reflected field when Bob's capture control is \emph{on} (blue dots on Fig.~3) or \emph{off} (red dots). The ratio of the reflected powers matches the one predicted through simulation (plain lines), validating our calibrations.\\

\section{Remote entanglement}\label{sec:entangle} We now turn to the task of entangling Alice and Bob. This is done by first having Alice release  ``half'' of a photon and thus getting entangled with the traveling mode in the state $(|g0\rangle+|e1\rangle)/\sqrt{2}$, which corresponds to the Hadamard and CNOT gates in Fig.~1a. This operation is followed by a \emph{swap} gate between the traveling mode and Bob, which corresponds to the same capture sequence as for the excitation transfer. The controls are determined with the same constraints but for the total integrated photon flux in the line being 1/2. The amplitude of $g_\mathrm{s}$ in this case is smaller than for the full release, so that we can use a traveling wavepacket with a reduced characteristic time $\sigma=450~\mathrm{ns}$. We plot the measured populations of Alice and Bob during the transfer on Fig.~4a (red and blue dots), which agree with the simulation predictions (lines) performed with the same parameters. We also plot the measured correlator $ \langle Z_{A}Z_{B}\rangle_{meas}$ (where $Z=2|e\rangle \langle e|-1$) between these measurements (green dots). When considering the correlations after correcting for readout errors (dashed lines), we find that at final time the actual occupation of the excited state is $P(|e\rangle_{A})=0.5$ and the actual correlator is $\langle Z_{A}Z_{B}\rangle = 2 P(|e\rangle_{B})$ (within 1~\%), which implies that Bob is excited \emph{only if} Alice is. In other words, as a photon detector, Bob's false positive probability beyond dispersive readout imperfections is below our detection precision (set by the accuracy of our simulations around 1~\%). This behavior is confirmed by the efficient cooling of the qubits when driving $|e0\rangle \leftrightarrow |g1\rangle$ prior to the experiment~\cite{supmat}. This property is crucial in non-deterministic entangling schemes, where the catch protocol could be used to perform single microwave photon detection~\cite{Narla2016,besse2017single, kono2017quantum}.\\
\begin{figure}[]
\includegraphics[scale=0.4]{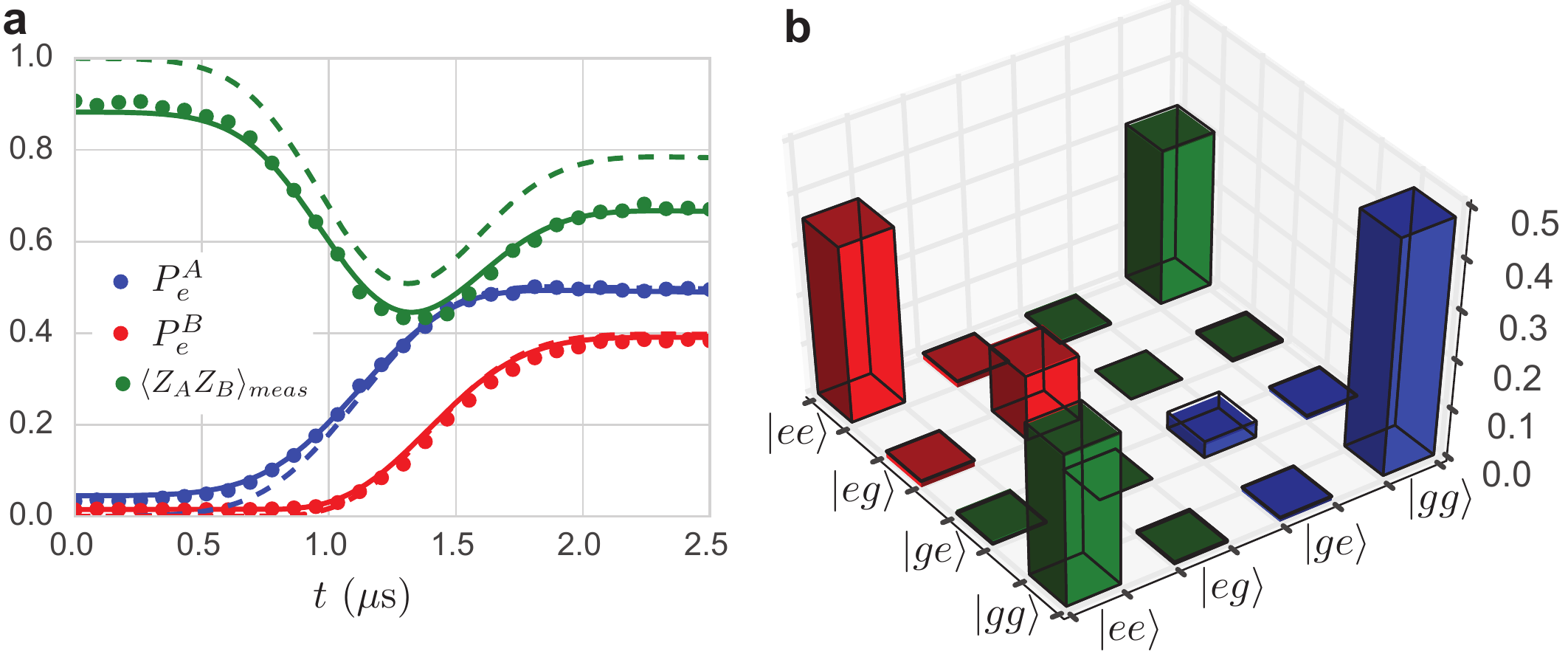}
\caption{\label{fig:half}  With Alice and Bob initially in $|g\rangle$, a  pump control signal is applied on Alice to release half a photon (see text) while the capture sequence of Fig.~\ref{fig:full}b is played for Bob. \textbf{a)} Measured excited state populations and correlator (with $Z=2|e\rangle \langle e|-1$)   when interrupting the control pulses after a duration $t$ and then performing simultaneous dispersive readout on both qubits. Plain lines are simulations including all imperfections. Dashed lines are the same simulations assuming perfect final readouts. \textbf{b)} Real part of the density matrix of the final entangled state measured by tomography of the two-qubit state (colored bars) and reconstituted by simulation (black contours). Fidelity to the Bell state $(|gg\rangle+|ee\rangle)/\sqrt{2}$ is 73~\%.
 }
\end{figure}
Finally, we perform full tomography of the final joint state of Alice and Bob. This is done by performing $\pi/2$ rotations of the qubits prior to readout to measure all Pauli operators $X, Y$ and $Z$ of each qubit and their correlators. After rotating the $(X_\mathrm{B}, Y_\mathrm{B})$ basis to compensate for the \emph{a priori} unknown but deterministic differential phase accumulated by control and pump pulses along the input lines~\footnote{For this measurement, the phases of the local oscillators used to generate the various pulses need to be locked in order to avoid drifts of this angle during averaging.}, one can  directly compute the density matrix following $\rho=\frac{1}{4}\sum_{\alpha, \beta \in \{I, X, Y, Z\}} \langle \alpha_{A}\beta_{B}\rangle_{meas}~   \alpha_A  \otimes \beta_B$. The fidelity to the target Bell state $|\Phi_+\rangle = (|gg\rangle+|ee\rangle)/\sqrt{2}$ is found to be $F=\mathrm{Tr}(\rho |\Phi_+\rangle \langle \Phi_+ |)=73~\%$, well above the entanglement threshold $F=1/2$. Once again, the measured density matrix (colorbars on Fig.~4b, see~\cite{supmat} for a full representation of the two-qubit state Pauli vector components) is in quantitative agreement with simulation predictions (black transparent bars).\\

In this experiment, we have implemented a simple protocol to perform reliable operations between standing qubits and arbitrarily shaped traveling photons. The method has been validated by fast (2.5~$\mu$s) remote entanglement of two qubits separated by $\sim1$~m microwave cables and a circulator. This protocol could be readily extended to entangle two pairs of qubits shared by Alice and Bob in order to correct for photon losses in the transmission line by entanglement purification~\cite{Bennett1995, Kalb2017}. Moreover, by controlling the traveling photon wavepacket shape in frequency, the signal from one cavity could be routed to another arbitrary one connected on the same line. All these features are important primitives on the path to a reliable modular quantum computing architecture~\cite{Monroe2014} or quantum internet~\cite{Kimble2008}.\\
The authors thank Z. Leghtas, A. Grimm and S. Touzard for helpful discussions, and M. Rooks for fabrication assistance. Facilities use was supported by the Yale Institute for Nanoscience and Quantum Engineering (YINQE), the National Science Foundation (NSF) MRSEC DMR 1119826, and the Yale School of Engineering and Applied Sciences cleanroom. This research was supported by the U.S. Army Research Office (Grant No. W911NF-14-1-0011), and the Multidisciplinary University Research Initiative through the U.S. Air Force Office of Scientific Research (Grant No. FP057123-C). L.B. acknowledges support of the ARO QuaCGR Fellowship.

\bibliography{bibliography2}{}
\bibliographystyle{unsrt}

\setcounter{equation}{0}
\setcounter{figure}{0}
\setcounter{table}{0}
\setcounter{page}{1}
\makeatletter
\renewcommand{\theequation}{S\arabic{equation}}
\renewcommand{\thefigure}{S\arabic{figure}}
\renewcommand{\bibnumfmt}[1]{[S#1]}
\renewcommand{\citenumfont}[1]{S#1}

\clearpage
\widetext
\begin{center}
\textbf{\large Supplemental Materials: Deterministic remote entanglement of superconducting circuits through microwave two-photon transitions}
\end{center}

\renewcommand{\thefigure}{S\arabic{figure}}

\makeatother

\section{Experimental setup and system characterization}

\begin{figure}[h!]
\includegraphics[scale=0.64]{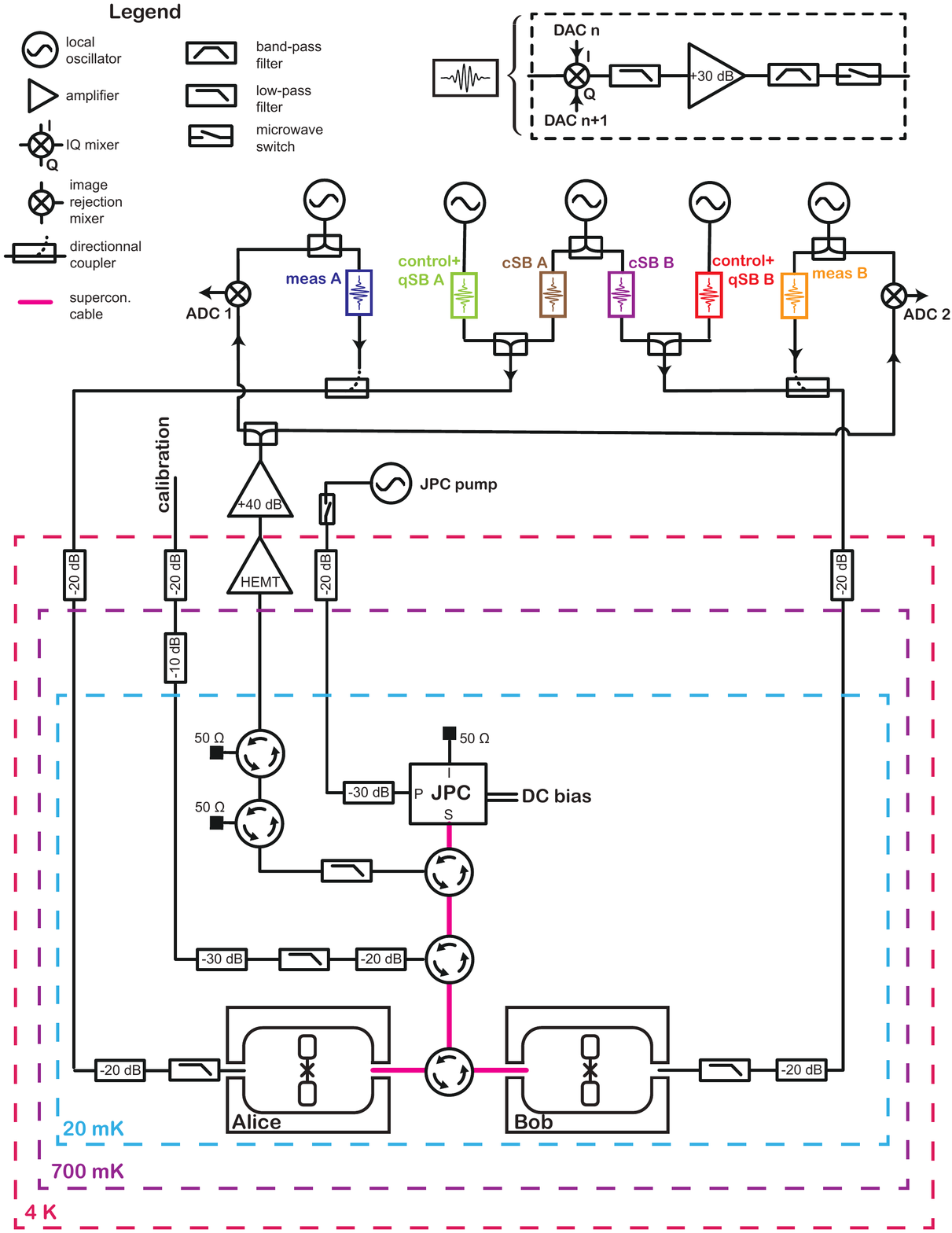}
	\caption{\label{fig:setupsup}  \textbf{Full wiring diagram} The two transmon-cavity systems are anchored on a dilution fridge base plate. Microwave pulses used to probe or control the system are generated by IQ mixing of a local oscillator (LO) with a low frequency signal with arbitrary envelope delivered by the DAC of an integrated FPGA system. These pulses include qubit rotation pulses (\emph{control}), qubit sideband pumps (\emph{qSB}), cavity side-band pumps (\emph{cSB}) and dispersive measurements (\emph{meas}). They propagate down heavily attenuated lines and drive  the two cavities  through their weakly coupled port. We use circulators and dissipationless superconducting cables to route the output field from Alice to Bob and then amplify it by reflection on the signal port (label S) of a pumped Josephson Parametric Converter used as a phase preserving amplifier (the idler port labeled I is fed with the vacuum from a thermalized 50~$\Omega$ load). The signal is then further amplified and down-converted at room-temperature using the same LO as for measurement pulses generation before digitization by the FPGA board. The \emph{calibration} line is used to measure the attenuation of the line between Alice and Bob. The room temperature setup is slightly modified when recording the traveling field presented on Fig.~3c of the main text in order to down-convert the signal with the LO used to generate cSB pulses.
}
\end{figure}
The transmons and cavities used in the experiment correspond to the qubits labeled \emph{Alice} and \emph{detector} in a previous publication by Narla \emph{et al.}~\cite{Narla2016}. The fabrication process is described in details there. The transmons are anchored in indium-plated copper cavities each probed through 2 ports. A coaxial coupler makes the weakly coupled one (photon exit rate $\kappa_{in}/2\pi\sim~5~\mathrm{kHz}$) while an aperture at the antinode of the TE101 mode opening to a waveguide-to-coaxial-cable adapter makes the strongly coupled one ($ \kappa/2\pi =1~\mathrm{MHz}$). Dissipation through this port dominates over cavity internal losses ($\kappa_L \leqslant \kappa_{in}$). The two cavities resonance frequency are finely tuned at room temperature by inserting into each an aluminum screw at the TE101 antinode. This allows us to match the resonance frequencies within 600~kHz at base temperature. The cavities are protected from external magnetic fields using  $\mu$-metal.\\

The full wiring diagram of our experiment is depicted on Fig.~S1. The microwave lines are filtered using both homemade \emph{eccosorb}-based dissipative filters and commercial reflective \emph{K}\&\emph{L} filters. The pulses used to probe and control the system are generated at room temperature by IQ mixing of a local oscillator provided by a microwave source at $\omega+\omega_h$ with a low frequency signal at $\omega_h$ ($ 30~\mathrm{MHz}< \omega_h/2\pi < 130~\mathrm{MHz}$) with arbitrary envelope delivered by the Digital to Analog Converter (DAC) of an integrated FPGA system. The resulting signal is amplified to get sufficient driving power for the qubit and cavity sideband pumps. In order to avoid feeding the system with unnecessary noise, the output signal from the amplifiers is tightly filtered (commercial bassband filters with width $\sim$100~MHz) and allowed in the refrigerator only when needed by closing microwave switches when generating a pulse.\\

The qubits population and coherence  decay times $T_1$ and $T_2$ are measured and presented on Table.~1. The unusually low $T_2$ of our qubits is an important limitation to the performance of both the excitation transfer and the remote entanglement presented in the main text. Note that the $T_2$'s are also slightly decreased when applying the sideband pumps to drive the two-photon transition (value in parenthesis on Table.~1 corresponding to $\xi_{1A}\simeq \xi_{1B}\simeq 0.11$). In practice, this limits the conversion and squeezing strengths that we could use and thus puts a lower bound on the protocol duration. This unexpected effect may originate from the low detuning regime we placed ourselves in ($\Delta < \chi_{qq}$) and will be the object of a forthcoming publication.

\section{Qubits initialization and readout}
The qubits are readout through standard dispersive measurement~\cite{blais2004cavity}. A $8~\mu\mathrm{s}$-long pulse for Alice ($4~\mu\mathrm{s}$ for Bob) at $\omega_c$ is applied on the cavity through the input port and the transmitted signal is amplified by a Josephson Parametric Converter used as a phase-preserving amplifier. After further amplification (see Fig.~S1) the signal is down-converted and digitized at room temperature. The integrated signal is then compared to a threshold to decide whether the qubit is in $|g\rangle$ or $|e\rangle$. Note that Alice resonance frequency (when the qubit is in $|g\rangle$) is detuned enough from Bob's that we can perform a readout on Alice \emph{independent} on Bob's state (see Fig.~S2) even though the probe field bounces off Bob's cavity. In practice, we readout both qubits simultaneously. The JPC maximum of amplification is chosen in between the two readout frequencies at $(\omega_{cA}+\omega_{cB})/2$ and the amplification bandwidth of 4MHz is sufficient to retain about 15~dB of amplification at both frequencies. The fidelities given in Table.~S1 are extracted from the contrast of Rabi oscillations (see Fig.~S2) and assume perfect initialization of the qubits in $|g\rangle$.\\

The qubits are initialized in $|g\rangle$ before every experiment using a non-destructive measurement and a fast feedback loop~\cite{Riste2012,  Campagne-Ibarcq2013}. This is needed as the important occupation of the $|e\rangle$ level at equilibrium ($p_{eq}=10~\%$ for Alice, $p_{eq}=4~\%$ for Bob) is far above the thermal distribution expected at 20~mK. Each qubit is measured repeatedly until it is found to be in $|g\rangle$. If the measurement indicates that the qubit is in $|e\rangle$, a control pulse corresponding to a $\pi$ rotation is triggered in order to shorten the total preparation time. Here, the readout parameters are finely tuned to empirically maximize the contrast of subsequently recorded Rabi oscillations. Compared to the final readout, the important feature is no longer the distinguishability  between $|g\rangle$ and $|e\rangle$ on a single-shot but rather to make sure that the qubit is in $|g\rangle$ at the end of the protocol. This is achieved by choosing a more stringent thresholding criterion (while keeping reasonable preparation time) and decreasing the readout pulse amplitude and duration to improve quantum non destructiveness.\\

\begin{figure}[h!]
\includegraphics[scale=0.55]{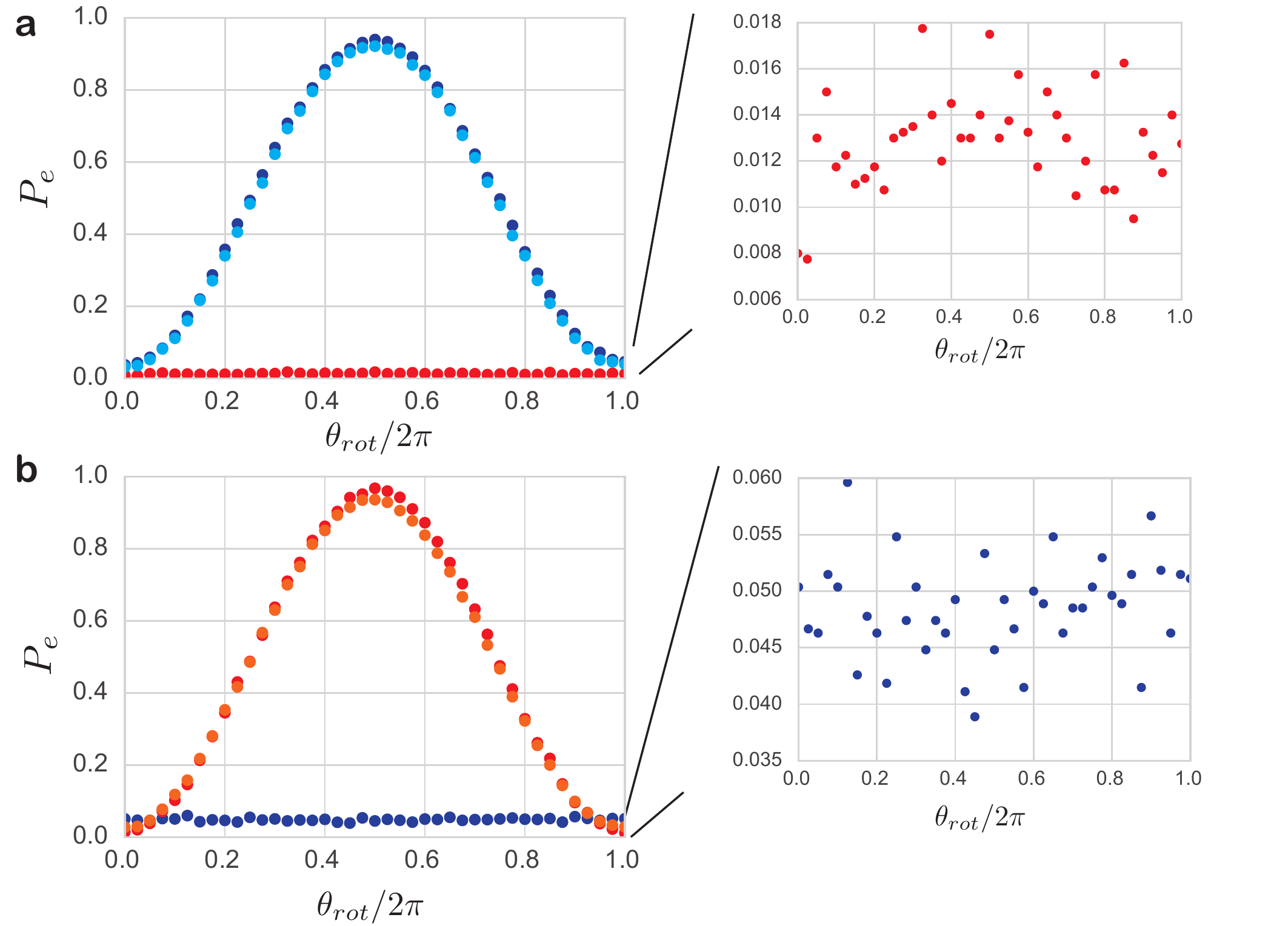}
\caption{ \textbf{Readout fidelities calibration} We record a Rabi oscillation of Alice (a) and Bob (b) in the excited state occupation of the qubits, when the qubit is initially prepared in $|g\rangle$ by measurement-based feedback only (light blue dots for Alice, orange dots for Bob) or by feedback followed by a 3~$\mu$s pulse on the two-photon transition $|e0\rangle \leftrightarrow|g1\rangle$ (dark blue dots for Alice, red dots for Bob). Assuming perfect preparation in this last case, the readout fidelities given on Table~I are extracted from the contrast of the oscillations. We also plot on the same graphics and on the magnified insets the detected population of Bob (red dots in (a)) and Alice (blue dots in (b)) when the other qubit is driven. No change in the qubit population is visible, demonstrating that the readouts are not correlated.
}
\end{figure}

In order to cool further the qubits, following the feedback loop, we apply sideband pumps in \emph{conversion} (see resonance condition~(3b) of the main text) for 3~$\mu$s with a power corresponding to a two-photon transition Rabi frequency of $g_c/2\pi\sim1-2~$MHz. Combined with the fast relaxation of the cavity photons into the lines ($\kappa \gg 1/T_1$), this drive brings the system to the dark state $|g0\rangle$. On Fig.~S2, the contrast of Rabi oscillations of the qubits is increased when using this scheme, showing that the qubits are indeed cooled in the process. The effect is more visible for Bob (Fig.~S2b) as for technical limitations of the FPGA, the feedback protocol described above is performed sequentially on Bob and \emph{then} on Alice so that Bob's qubit can re-excite during Alice's preparation.\\

This autonomous scheme for qubit initialization may prove useful in circuit-QED experiments as it does not rely on different cavity photon number giving resolved qubit transition peaks. Indeed, $|g0\rangle$ is still a dark state even when $\kappa_c> \chi_{cq}$. This is in contrast with other autonomous schemes such as the DDROP protocol~\cite{geerlings2013demonstrating}. Moreover, the fact that the qubit gets cooled to a point where the only events when it is detected in $|e\rangle$ can be attributed to readout errors gives a strong indication that, seen as a photon detector, the \emph{catch} gate has negligible false positive probability beyond these readout errors. Note that cooling is also observed when applying the non-constant pulse corresponding to a \emph{catch} by Bob when no photon is emitted by Alice.

\section{Transmission channel characterization}
\begin{figure}[]
\includegraphics[scale=0.6]{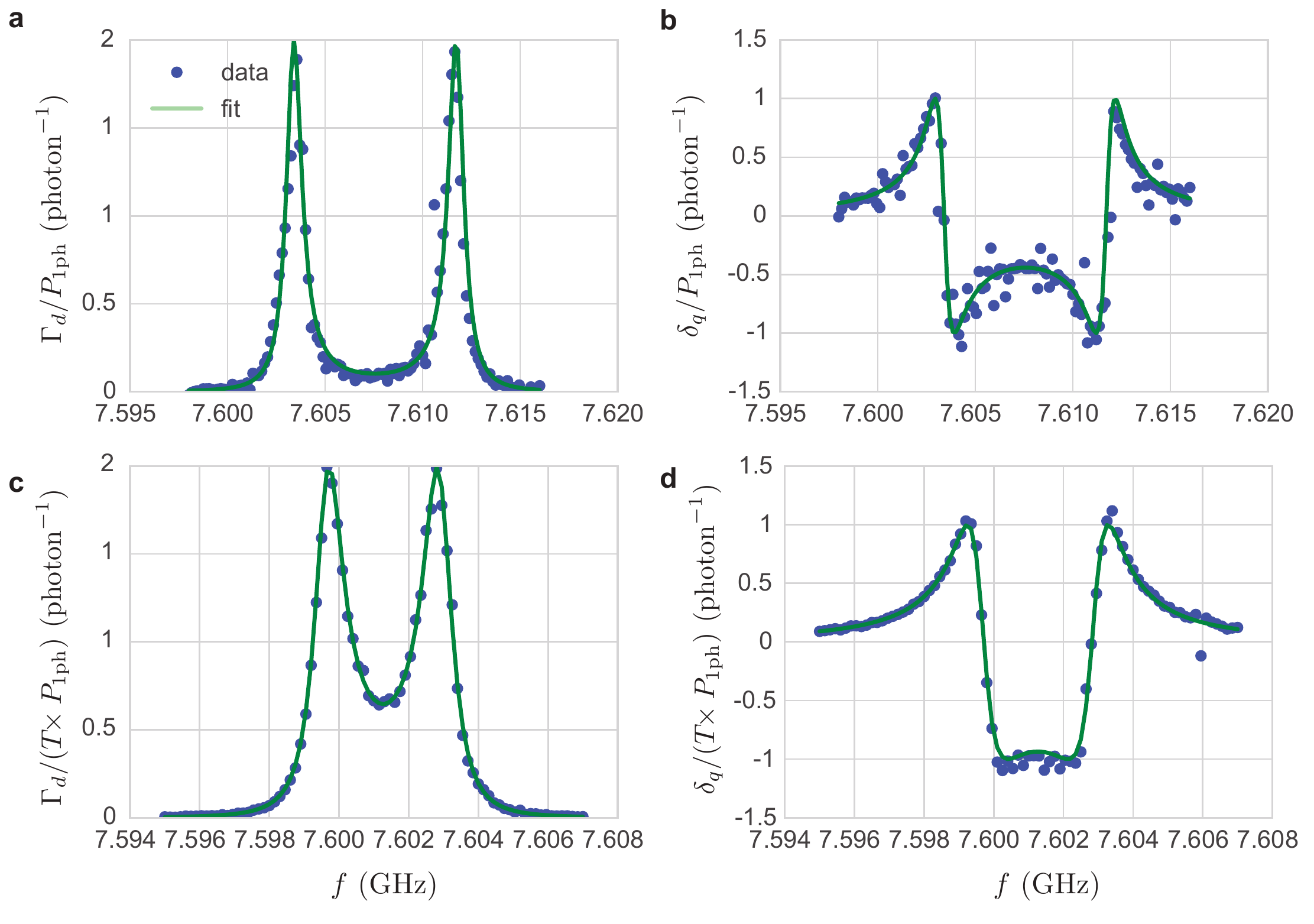}
\caption{\label{fig:MID}  \textbf{Dressed Ramsey interferences}  We plot the normalized value of the measurement induced dephasing $\Gamma_d$ and Stark shift $\delta_q$ for Alice ((a) and (b)) and Bob ((c) and (d)) when varying the frequency $\omega$ of a continuous wave bouncing off each cavity. The normalizing power $P_{1\mathrm{ph}}$ for Alice (corresponding to an average photon flux of 1~photon per unit time) and $T\times P_{1\mathrm{ph}}$ for Bob ($T$~photon per unit frequency) are extracted at the same time as the values of $\kappa_c$ and $\chi_{cq}$ when fitting each pair of curves (green lines). The transmission of the line is found to be $-0.7~\mathrm{dB}$.
  }
\end{figure}

In order to calibrate the probability of photon loss in the microwave line and circulator between Alice and Bob, we use the \emph{calibration} line (see Fig.~S1). We apply a continuous wave tone around the cavity resonance frequencies through this port, which bounces off Alice's cavity (input power $P_A$), and then Bob's (input power $P_B$). For a given microwave power at the refrigerator input, $P_A$ and $P_B$ are \emph{a priori} unknown due to the uncalibrated attenuation of the lines. The ratio $T=P_B/P_A$ is precisely the transmission of the line that we want to calibrate. We then perform a Ramsey interference experiment on Alice and then Bob in presence of this dressing field. Following Gambetta \emph{et al.}~\cite{gambetta2008quantum}, the oscillation and decay of the Ramsey fringes should reveal the measurement induced dephasing $\Gamma_d$ and Stark shift $\delta_q$ of each qubit by the dressing field as
\begin{equation}
\begin{split}
\Gamma_d=& \chi_{cq} \mathrm{Im}(\alpha_e \alpha_g^{\ast})\\
\delta_q=& \chi_{cq} \mathrm{Re}(\alpha_e \alpha_g^{\ast})
\label{midd}
\end{split}
\end{equation}
where $\mathrm{Re}()$ and $\mathrm{Im}()$ denote the real and imaginary parts and $\alpha_e=\frac{2\sqrt{\kappa P}}{\kappa+2 i (\omega-\omega_c+\chi_{cq})}$ is the steady state of the intra cavity field under the continuous driving at $\omega$ with microwave power $P=P_A, P_B$ when the qubit is in $|e\rangle$, and $\alpha_g=\frac{2\sqrt{\kappa P}}{\kappa+2 i (\omega-\omega_c)}$ when the qubit is in $|g\rangle$.\\
 If, for Alice and Bob, the cavities photon exit rates $\kappa$ and dispersive shifts $\chi_{cq}$ were independently calibrated, one could directly access the values of $P_A$ and $P_B$, and then $T$. In practice these two parameters are difficult to determine in a separate experiment. Alternatively, we vary the dephasing microwave frequency and extract the value of $\Gamma_d$ and $\delta_q$ for each value of $\omega$. For better precision, we actually also vary the power $P_0$ of the dephasing microwave at the refrigerator input and consider  the slope of $\Gamma_d$ and $\delta_q$ against $P_0$. For each qubit, we then fit both curves $\frac{\Gamma_d}{P_0}(\omega)$ and $\frac{\delta_q}{P_0}(\omega)$ with three parameters $\chi_{cq}$, $\kappa$ and $P/P_0$. The fitted data is represented on Fig.~S3, where we have scaled Alice's dephasing and Stark shift to correspond to the ones induced by a flux of 1~photon per unit time (Bob scaling corresponds to T~photon per unit time). The extracted value of the line transmission is $T=0.85~\pm0.05$. This relatively large uncertainty originates from low frequency noise on the qubits resonance frequency (particularly visible on Alice on Fig.~S3b) and deviation of the measured $\Gamma_d$ and $\delta_q$ with respect to the model Eq.~\eqref{midd}, probably due to imperfect isolation by the circulator. The value of $T$ is however confirmed by the quantitative agreement between the recorded traveling wavepacket  and  predictions of simulations with this value of $T$  when pitching a photon but not catching it (see Fig.~3c). Not that this value corresponds to 0.7~dB attenuation,  above the nominal insertion loss of our circulator (0.2~dB) if we assume that the superconducting line is dissipationless.

\begin{center}
\begin{table}[!h!t!b!p]
\begin{tabular} {|c||c|c|}
\hline
~&~~~~~~Alice~~~~~~&~~~~~~Bob~~~~~~\\
\hline

$T_1~(\mu\mathrm{s})$& 100 & 96 \\ 
\hline
$T_2~(\mu\mathrm{s})$& 11.5 (10) & 20 (17.5) \\ 
\hline

$\omega_c /2\pi~(\mathrm{GHz})$& 7.6118 & 7.6029 \\ 
\hline
$\omega_q /2\pi~(\mathrm{GHz})$& 4.510 & 4.751 \\ 
\hline
$\kappa/2\pi~(\mathrm{MHz})$& 1.0 & 1.0 \\ 
\hline
$\chi_{cq}/2\pi~(\mathrm{MHz})$& 8.3 & 3.3 \\ 
\hline
$\chi_{qq}/2\pi~(\mathrm{MHz})$& 200 & 240 \\ 
\hline
$\chi_{cc}/2\pi~(\mathrm{kHz})$& 85 & 10 \\ 
\hline
$~~~|g\rangle$ readout fidelity~~~ & 0.955 & 0.985 \\ 
\hline
$~~~|e\rangle$ readout fidelity~~~ & 0.94 & 0.96 \\ 
\hline
\end{tabular}
\caption{  Alice and Bob parameters and experimental imperfections. $T_2$ value between parenthesis is with the qubit sideband pump on.
}
\end{table}
\end{center}
\section{Driven non-linearity : resonant processes and pumps calibration}

\subsection{Hamiltonian derivation}
\label{sec:hamil}
We consider one transmon in a cavity. Following Nigg \emph{et. al.}~\cite{nigg2012black}, one can formally split the Josepshon junction into a linear inductor and a purely non-linear element. The environment seen by this non linear element is a series of coupled linear modes with low dissipation (the plasma excitations of the junction shunted by the antennas and the modes of the cavity). One can then find a decoupled mode basis (Foster decomposition) of this environment, whose two first resonant modes are labeled $q$ and $c$ and correspond to the transmon and cavity modes considered in the main text. We now neglect higher frequency modes. In presence of the two pumps ($j=1,2$), which each drive both the transmon and the cavity mode ($a=q, c$) with strengths $\epsilon_{aj}$, the Hamiltonian of the system reads
\begin{equation}
\begin{split}
\frac{H}{\hbar}=&\omega_q q^{\dagger}q+\omega_c c^{\dagger}c -\frac{E_J}{\hbar}~(\mathrm{cos}(\varphi) + \frac{\varphi^2}{2})+\mathop{\mathop{{\sum}}_{a=q,c}}_{ j=1,2} 2\mathrm{Re}(\epsilon_{aj}e^{-i\omega_i t})(a+a^{\dagger})
\label{hamilsup1}
\end{split}
\end{equation}
where $\mathrm{Re}()$ denotes the real part, and the phase across the non linear element $\varphi$ is the sum of the contributions from the transmon and cavity modes (zero point fluctuations $\varphi_q$ and $\varphi_m$)
\begin{equation}
\varphi=\varphi_q (q+q^{\dagger}) + \varphi_c (c+c^{\dagger})
\end{equation}
Then, following Leghtas \emph{et. al.} (see supplementary material of \cite{leghtas2015confining}), we move to a four times displaced frame with the unitary
\begin{equation}
U=\mathop{\mathop{{\prod}}_{a=q,c}}_{ j=1,2} e^{-\tilde{\xi}_{aj} a^{\dagger}+\tilde{\xi}^{\ast}_{aj}a}=e^{i(\theta_q+\theta_c)}e^{  \mathop{\mathop{{\sum}}_{a}}_{ j}    -\tilde{\xi}_{aj} a^{\dagger}+\tilde{\xi}^{\ast}_{aj}a }
\end{equation}
where the phases $\theta_{q,c}$ resulting from the non commutation of $\tilde{\xi}_{ai,aj} a^{\dagger}+\tilde{\xi}^{\ast}_{ai,aj}a$ when $i\neq j$ give a global phase not physically relevant. The displacements are chosen to be $\tilde{\xi}_{aj}=\xi_{aj} e^{-i\omega_jt}$ with $ \xi_{aj}=\frac{\epsilon_{aj}}{\frac{\kappa_a}{2}+i(\omega_a-\omega_j)}$. Here, $\kappa_a$ is the dissipation rate of mode $a$ and can be neglected in the transmon case. Note that each displacement corresponds to the steady state amplitude of the considered mode when driven continuously by the pump. \\

We place ourselves in a regime where $\omega_{q,c}\gg |\Delta|=|\omega_q-\omega_1|=|\omega_c-\omega_2|\gg |\xi_{1q,1c,2q,2c}|^2 E_J\varphi_{q,c}^4$. This allows us to simplify the dynamics of the system in the displaced frame by neglecting fast rotating terms. The evolution of the system state then appears to be governed by the effective Hamiltonian
\begin{equation}
\begin{split}
&~~\frac{\tilde{H}}{\hbar}=\omega_q \tilde{q}^{\dagger}\tilde{q}+\omega_c \tilde{c}^{\dagger}\tilde{c}-\frac{E_J}{\hbar}(\mathrm{cos}(\tilde{\varphi}) + \frac{\tilde{\varphi}^2}{2})\\
&\tilde{\varphi}=\varphi_q (\tilde{q}+\tilde{q}^{\dagger}-\tilde{\xi}_1-\tilde{\xi}^{\ast}_1) + \varphi_c (\tilde{c}+\tilde{c}^{\dagger}-\tilde{\xi}_2-\tilde{\xi}^{\ast}_2)\\
&~~~~~~~~~~\tilde{q}=q+\tilde{\xi_1}~~,~~ \tilde{c}=c+\tilde{\xi_2}
\end{split}
\end{equation}
where we have written the total displacement, shared by both modes, due to each pump as
\begin{equation}
\begin{split}
 \tilde{\xi}_{1}=\xi_1e^{-i\omega_1t}=(\xi_{1q}+\frac{\varphi_c}{\varphi_q}\xi_{1c})e^{-i\omega_1t}\\
 \tilde{\xi}_{2}=\xi_2e^{-i\omega_2t}=(\xi_{2c}+\frac{\varphi_q}{\varphi_c}\xi_{2q})e^{-i\omega_2t}
 \label{eq:xissup}
 \end{split}
\end{equation}
Thus, the situation is equivalent to the simplified case where we consider that each pump addresses only one mode except that the proportionality factor between the drive strength $\epsilon_j$ and the effective displacement $\xi_j$ is \emph{a priori} unknown and can vary with the pump frequencies. This can be seen as an interference effect between two paths, through mode $q$ and mode $c$, for the pump microwave to reach the junction.\\

Developing this Hamiltonian to the fourth order (we are in the regime $|\tilde{\varphi}|\ll 1$) and considering only terms that may be resonant, the Hamiltonian takes the form (1) given in the main text
\begin{equation}
\begin{split}
\frac{H}{\hbar}=~&(\omega_q+ \delta_q) q^{\dagger}q+(\omega_c+ \delta_c) c^{\dagger}c -\frac{\chi_{qq}}{2}(q^{\dagger}q)^2-\frac{\chi_{cc}}{2}(c^{\dagger}c)^2 -\chi_{cq} q^{\dagger}q c^{\dagger}c\\
&~~~+e^{-i (\omega_1+\omega_2)  t} g_{\mathrm{s}}~q^{\dagger} c^{\dagger} + h.c. \\
&~~~+e^{-i (\omega_1-\omega_2) t} g_{\mathrm{c}}~q^{\dagger} c + h.c.
\label{hamilsup2}
\end{split}
\end{equation}
The Kerr terms are $\chi_{cc,qq}=E_J\frac{\varphi_{c,q}^4}{2}$ and  $\chi_{cq}=E_J\varphi_c^2 \varphi_q^2$ while the Stark-shifts, squeezing and conversion terms read~\footnote{The bare frequencies $\omega_q$ and $\omega_c$ are also shifted by a constant value when commuting the operation and annihilation operators to get the normal ordering of Eq.~S7}
\begin{equation}
\begin{split}
\delta_q=&~  -2\chi_{qq} |\xi_1|^2 -\chi_{cq} |\xi_2|^2 \\
\delta_c=&~  -2\chi_{qq} |\xi_1|^2 -\chi_{cq} |\xi_2|^2  \\
g_{\mathrm{s}}=&~-\chi_{cq} \xi_1 \xi_2 \\ 
g_{\mathrm{c}}=&~-\chi_{cq} \xi_1 \xi_2^{\ast} 
\label{eq:starkssup}
\end{split}
\end{equation}
(the minus signs of the last two expressions has been absorbed in the pump phases in the main text). \\

From these expressions, one can estimate the maximum cavity frequency variation when modulating $\xi_2$ to get the desired values of $g_c$ and $g_s$ in the experiment (see Fig.~3a-b of the main text). With $\chi_{ccA}=85~\mathrm{kHz}$ and $\chi_{ccB}=10~\mathrm{kHz}$ calculated from the Kerr expression given above, we get a maximum deviation of 70~kHz for Alice and 15~kHz for Bob. These detunings remain small compared to cavity linewidth $\kappa_c$ and $1/T_{on}$, where $T_{on}$ is the duration during which $\xi_2$ takes a significant value. This justifies the hypothesis of negligible cavity anharmonicity used to get to the expression (2) of the main text.\\

Note that all operators considered in the text are displaced when the pumps are \emph{on}. The pump amplitudes are made to vary adiabatically (on a time scale long compared to $1/\Delta$), so that the rotating frame approximations used above remain valid. The  frame displacement preserves the form of the dissipation so that the input-output relation $c_{\mathrm{A}}=1/\sqrt{\kappa}~a_{\mathrm{t}}=-c_{\mathrm{B}}$ given in the main text is valid, but the traveling mode $a_{\mathrm{t}}$ is also displaced. However, when demodulating at $\omega_c$ and recording the mean value of the field as presented on Fig.~3c of the main text, the signal is analogically filtered before digitization (low-pass filter with cut-off at 30~MHz) so that the fast oscillating component at $\Delta=100~\mathrm{MHz}$ corresponding to this displacement is filtered out.

\subsection{Stark shift measurements}
\label{sec:stark}
\begin{figure}[]
\includegraphics[scale=0.6]{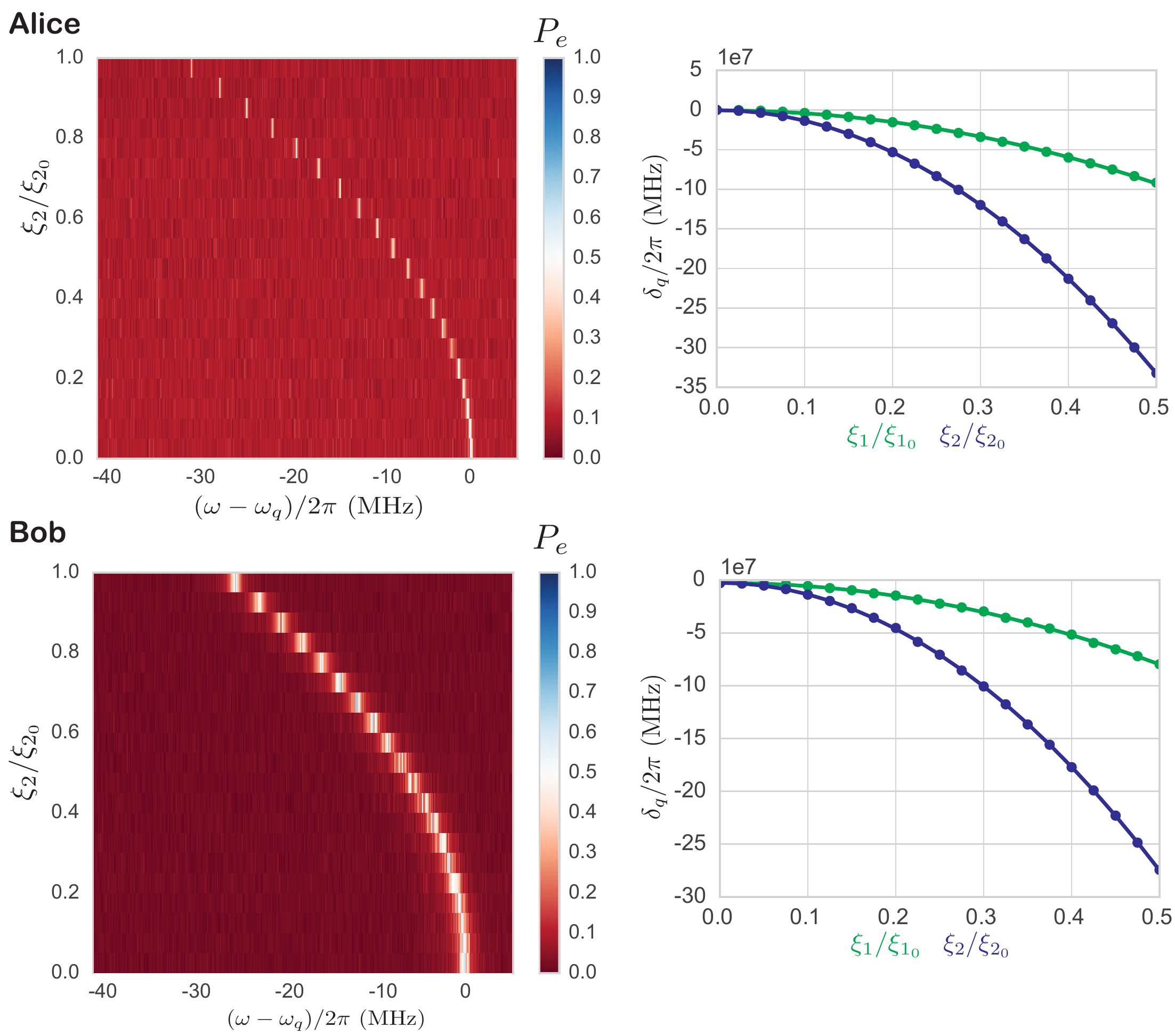}
\caption{\label{fig:stark} \textbf{Dressed qubit spectroscopy} We apply on Alice or Bob a 14~$\mu$s-long cavity sideband pump (qubit sideband pump) of varying amplitude $\xi_2$ ($\xi_1$) while probing the qubit with a low power microwave around its resonance frequency $\omega_q$. As can be seen on the left color plots  (recorded when varying  $\xi_2$ while $\xi_1=0$), a subsequent readout of the qubit reveals a desaturated line whose position varies with the sideband pump amplitudes. We extract the line position as a function of $\xi_2$ ($\xi_1$) and plot it as blue dots (green dots) on the right panel. The scaling factors $\xi_{1_0}$  and $\xi_{2_0}$ correspond to the largest amplitudes available in our setup. Quadratic fits following Eq.~\eqref{eq:starkssup} give $\xi_{1_0}=0.15$ and $\xi_{2_0}=2.0$ for Alice, $\xi_{1_0}=0.13$ and $\xi_{2_0}=2.9$ for Bob (lines on the right panels).
}
\end{figure}

\begin{figure}[]
\includegraphics[scale=0.6]{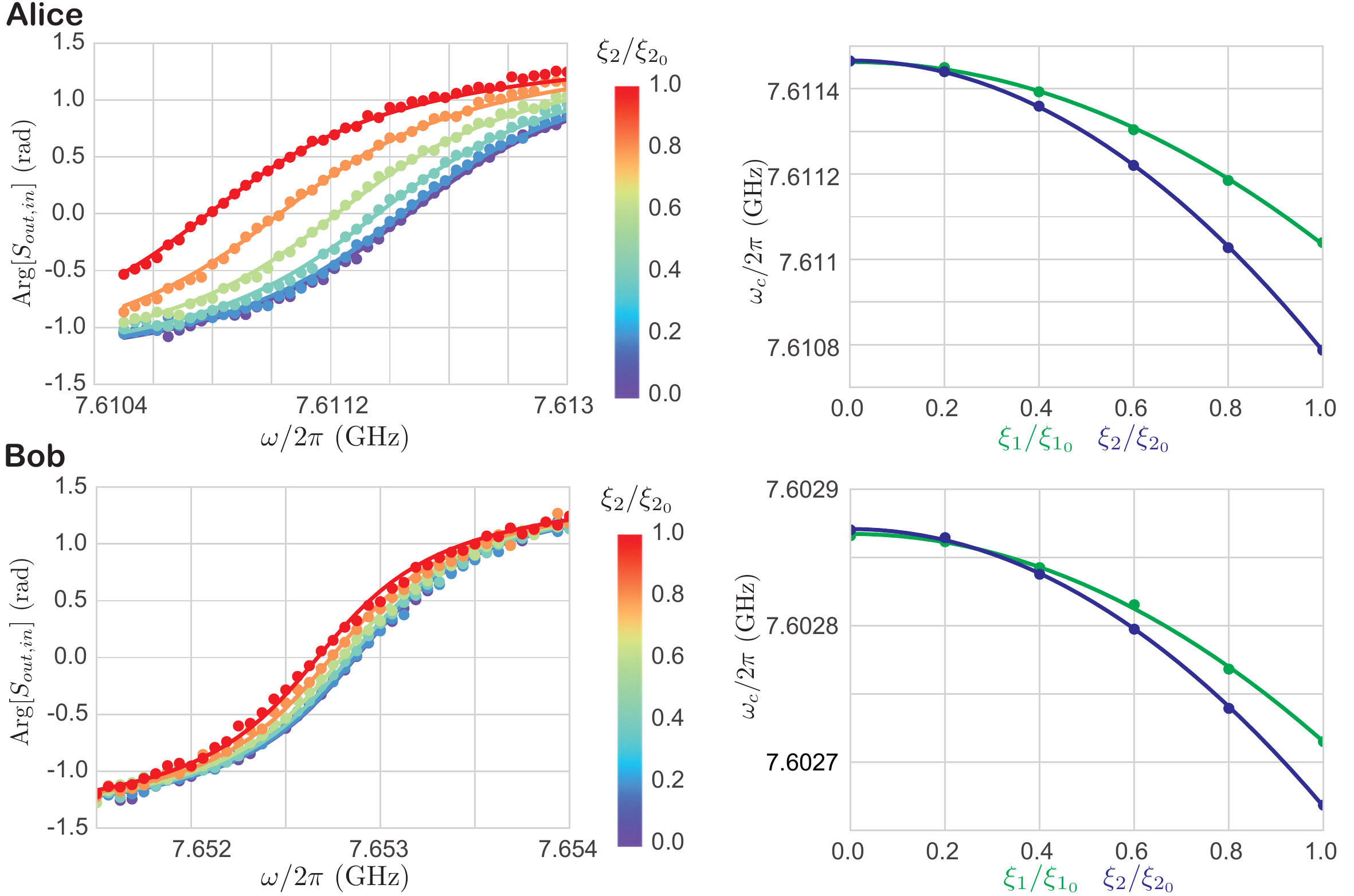}
\caption{\label{fig:stark2} \textbf{Dressed cavity spectroscopy} We apply on Alice or Bob a 14~$\mu$s-long cavity sideband pump (qubit sideband pump) of varying amplitude $\xi_2$ ($\xi_1$) while probing the cavity with a low power microwave around its resonance frequency $\omega_c$. After unraveling the transmitted signal to compensate for the propagation delay in the lines, the transmitted field displays a characteristic $\pi$ phase shift across resonance (dots on the left panels when varying $\xi_2$ and setting $\xi_1=0$). We fit this data with the resonance frequency as the only free parameter (lines on the left panel). The extracted value is plotted against $\xi_2$ ($\xi_1$) as blue dots (green dots) on the right panel. The scaling factors $\xi_{1_0}$  and $\xi_{2_0}$ correspond to the largest amplitudes available in our setup. Quadratic fits following Eq.~\eqref{eq:starkssup} give $\xi_{1_0}=0.23$ and $\xi_{2_0}=2.0$ for Alice, $\xi_{1_0}=0.22$ and $\xi_{2_0}=3.1$ for Bob (lines on the right panels).
 }
\end{figure}

We now turn to calibrating the \emph{a priori} unknown scaling factors linking the amplitude of the sideband pumps applied on Alice or Bob to the effective displacements $\xi_1$ and $\xi_2$ of the qubit and cavity modes (see Eq.~\eqref{eq:starkssup}). For each qubit, we vary the amplitudes of these applied drives and  measure the Stark shift $\delta_c$ and $\delta_q$ of the cavity and qubit modes, whose theoretical expressions are given in Eq.~\eqref{eq:starkssup}, in separated two-tone spectroscopy experiments. For clarity, we now describe these measurements for Alice only.\\

In two separate experiments, we apply on Alice either a 14-$\mu$s qubit  sideband  pulse of amplitude $\xi_1/\xi_{1_0}$ or a cavity sideband pulse of amplitude $\xi_2/\xi_{2_0}$, while simultaneously driving the qubit with a low power tone at $\omega$
 . Here $\xi_{1_0}$ and $\xi_{2_0}$ are the maximum displacements that we can achieve in our setup (corresponding to 0.5~V applied on the IQ mixers of Fig.~S1) and are precisely the values that we want to calibrate.  When varying $\omega$ around the bare qubit frequency $\omega_q$ and subsequently measuring the qubit excited state probability, we find a desaturated line at $\tilde{\omega}_q=\omega_q+\delta_q$. This is represented in color on Fig.~S4a when varying the cavity sideband pump amplitude $\xi_2/\xi_{2_0}$. We plot on the right panel the extracted value of $\delta_q$ for each sideband pump amplitude for the experiment in which we vary $\xi_2/\xi_{2_0}$ (blue dots, $\xi_1=0$) and for the other in which we vary $\xi_1/\xi_{1_0}$ (green dots, $\xi_2=0$). We observe the expected quadratic dependence  and fit this data for the value of $\xi_{1_0}=0.15  $ and $\xi_{2_0}=1.99$.\\
 
It is interesting to compare these values to the ones extracted from a second type of experiment, in which we perform a cavity spectroscopy and consider the cavity Stark shift $\delta_c$. The measurement consists in applying a 14~$\mu$s-long sideband pump (cavity \emph{or} qubit)  on the system while directly recording the transmitted amplitude of a low power probe around the  cavity resonance frequency $\omega_c$ (see Fig.~S5 left panels). We then fit the cavity $180^\circ$ phase shift across resonance with a lorentzian shape in order to extract the dressed resonance frequency $\omega_c+\delta_c$. This value is plotted against $\xi_{1}/\xi_{1_0}$ and $\xi_{2}/\xi_{2_0}$ (green and blue dots on the right panel on  Fig.~S5) and again shows a quadratic dependence on these two values. When fitting this data, we get $\xi_{1_0}=0.23$ and $\xi_{2_0}=1.98$. Thus, while the values of $\xi_{2_0}$ extracted from qubit and cavity spectroscopy measurements agree quantitatively,  those of $\xi_{1_0}$ differ by about 50~\%. A similar discrepancy is observed for Bob (see captions of Fig.~S4 and Fig.~S5).\\

When measuring the two-photon Rabi oscillation presented on Fig.~2 and Fig.~S7, the dependence of the Rabi frequency against the cavity sideband pump amplitude is well predicted using the value of $\xi_{1_0}$  extracted from cavity spectroscopy measurements (black dashed line) and not with the one from qubit spectroscopy measurements. The values of $\xi_{1A}$ and $\xi_{1B}$ given in the main text and in the next sections of the \emph{Supplementary Materials} are thus based on the Rabi oscillations and cavity spectroscopy calibration. \\

The model of Eq.~\eqref{eq:starkssup} for the qubit Stark shift probably needs an important correction, whose origin and value is for now not understood. It may originate from the non confining nature of the transmon \emph{cosine} potential, as tunneling between wells of this potential can renormalize the expressions of Eq.~\eqref{eq:starkssup}. Note that this discrepancy does not affect our capacity to drive resonantly the two-photon transitions as $\xi_1$ is fixed in the experiment. Moreover, if one needs to vary this control, an empirical resonance condition as a function of $\xi_1$ and $\xi_2$ could be measured in replacement of Eq.~(3) of the main text. 


\subsection{Resonance matching}

In the experiment, we keep $\xi_1$ constant for Alice and Bob so that the cavity dressed frequencies $\tilde{\omega}_{cA,cB}$ are constant when neglecting the anharmonicity of the cavity modes (see Eq.~\eqref{eq:starkssup}). Modulation of the squeezing and conversion strengths $g_s$ and $g_c$ is realized by varying the cavity sideband pump amplitude $\xi_2$ only. When doing so, the qubit dressed frequency $\tilde{\omega}_q$ varies and we want to vary accordingly the frequency $\omega_2$ of the pump so that the resonance condition given by of Eq.~(3) of the main text is met. However, changing $\omega_2$ in turns modifies the effective amplitude $\xi_2$. This effect has two origins. First, fast control on $\omega_2$ is achieved by varying the mixing heterodyne frequency  applied on the IQ mixers (see Fig.~S1). A low pass filter with cutoff at 150~MHz is placed at the output of the FPGA Digital to Analog Converter (see Fig.~S1) so that the power of the generated microwave pulses drops when one changes this heterodyne frequency by a significant fraction of 150~MHz. This results in an apparent dispersion relation with strong variations for the lines. Note however that this dispersion function tends to be real, which means that the amplitude of $\xi_2$ varies but not its phase. Second, as mentioned in section \emph{Hamiltonian derivation}, the scaling factor linking the sideband pump amplitude to the effective displacement $\xi_2$ varies with the sideband detuning $\Delta$, potentially in a complex way.\\

In order to calibrate these effects, we use the qubits as effective spectrometers. We perform on Alice and Bob a dressed spectroscopy experiment similar to the one described in Fig.~S4. The qubit is probed with a weak microwave around $\omega_q$ while applying a cavity sideband drive generated with a constant amplitude (at the DAC level) but varying its frequency $\omega_{c\mathrm{SB}}$ around $\omega_2$. The result, presented on Fig.~S6a-b is a moving qubit line that reveals the spurious variation of $\xi_2$. For Bob, the  Stark shift $\delta_q$ decreases with $\omega_{c\mathrm{SB}}-\omega_2$ (Fig.~S6b) as the FPGA mixing frequency is positive : $\omega_2=\omega_{LO}+\omega_h$ with $\omega_h/2\pi=130~$MHz. Increasing $\omega_h$ to increase $\omega_{c\mathrm{SB}}$ then brings it closer to the low pass filter cutoff at 150~MHz so that the power of the pump at the mixer output decreases. For Alice, $\omega_h/2\pi=-70$~MHz is negative so that the opposite tendency is observed.\\ 

After extracting the position of the qubit line as a function of $\omega_2$ and interpolating the resulting data, one can compensate for the real part of this effective dispersion relation by adjusting the pump amplitude. All sideband pulses in the main text include both a frequency modulation to match the resonance condition and an amplitude modulation to compensate for this dispersion. In particular, a shaped  pulse as on Fig.~3 and Fig.~4 results in a chirped pulse (to preserve the resonance condition at each time) which itself requires to adjust the pump amplitude along the pulse duration.\\

\begin{figure}[]
\includegraphics[scale=0.6]{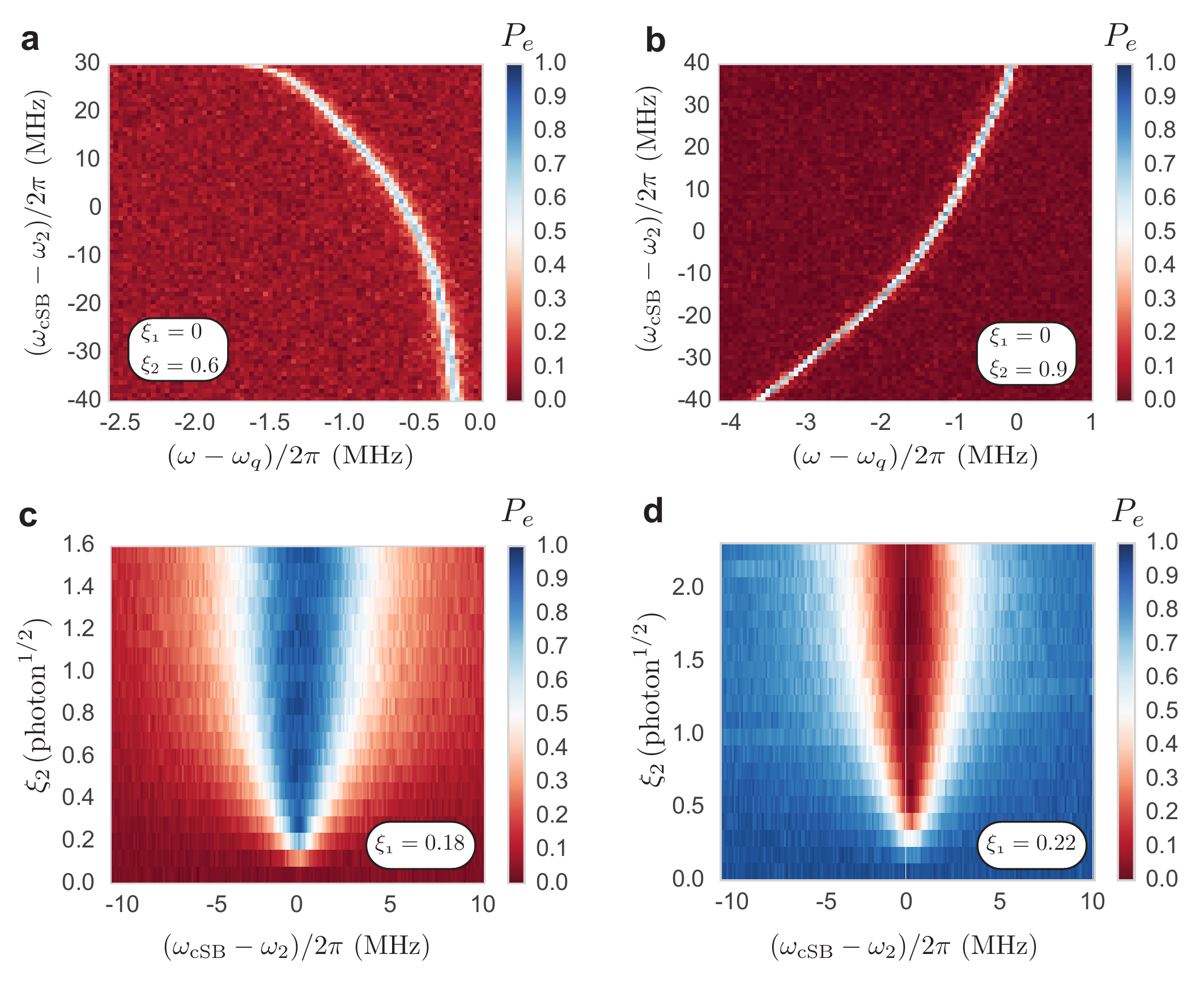}
\caption{\label{fig:sbspectro} Qubit dressed spectroscopy for Alice \textbf{(a)} and Bob \textbf{(b)} in presence of a 14~$\mu$s-long cavity sideband drive generated with a constant amplitude (at the DAC level) and varying frequency $\omega_{c\mathrm{SB}}$ around $\omega_2$ (y-axis of the color plot).  The drift of the qubit line when varying $\omega_{c\mathrm{SB}}$  reveals the effective dispersion relation of the line for the cavity sideband pump (see text). The given value of $\xi_2$ corresponds to the value calibrated in Fig.~S4 and Fig.~S5 at $\omega_{c\mathrm{SB}}=\omega_2$.  \textbf{(c)} With the line dispersion compensation \emph{on}, we perform a two-tone spectroscopy : for 3~$\mu$s, the qubit sideband pump amplitude $\xi_1$ and frequency $\omega_1$ are fixed, and we vary the  the cavity sideband pump amplitude $\xi_2$ (y-axis) and frequency $\omega_{c\mathrm{SB}}$ around $\omega_2(\xi_2)$ (x-axis). The qubit is then measured. A peak centered on $\omega_2$ appears and broadens when $\xi_2$ increases, demonstrating that the resonance condition is met at $\omega_{c\mathrm{SB}}=\omega_2$ for all $\xi_2$. \textbf{(d)} Same experiment for Bob, whose qubit is initialized in $|e\rangle$ before the spectroscopy.
  }
\end{figure}

To test our method, we perform a different spectroscopy experiment while using this compensation. After preparing Alice in $|g\rangle$, we apply  for 3~$\mu$s simultaneously a qubit sideband pump at $\omega_1$ corresponding to a fixed displacement $\xi_1$ and a cavity sideband pump at $\omega_{c\mathrm{SB}}$ around the resonant frequency $\omega_2$ for a given  displacement $\xi_2$. When sweeping $\omega_{c\mathrm{SB}}$ at $\xi_2=0$, a subsequent qubit readout gives a flat response (bottom of the color plot on Fig.~S6c) as the qubit sideband pump by itself does not change the qubit population. However, when increasing $\xi_2$ (and thus the central frequency $\omega_2$), we observe a rising and broadening peak always centered on the expected $\omega_2$. This sanity check confirms that we accurately meet the resonance condition $\tilde{\omega}_q+\tilde{\omega}_c=\omega_1+\omega_2$ when varying the two-photon drive amplitude. A similar experiment is performed on Bob, which is initially prepared in $|e\rangle$ (see Fig.~S6d). \\

On Alice, a residual drift of the peak is observed when increasing $\xi_2$. The observed detuning at $\xi_2=1.6$ is compatible with the expected cavity Stark shift due to its anharmonicity $2\chi_{ccA}|\xi_2|^2/2\pi\simeq0.4~$MHz (see  Eq.~\eqref{eq:starkssup}) that was neglected up to now. For Bob, who as a smaller anharmonicity, the effect is barely visible.\\

These spectroscopy measurements, combined with the Rabi oscillations presented on Fig.~2 and Fig.~S7 (which are their time domain counterparts), demonstrate our ability to drive one of the two-photon transition on resonance with a controlled strength $g_s$ or $g_c$.

\subsection{Two-photon Rabi oscillations}
In this section, we present two-photon Rabi oscillations recorded at smaller pump amplitudes compared to Fig.~2 of the main paper. The qubit sideband pump amplitude used here for Alice and Bob is $\xi_{1A}=\xi_{1B}=0.11$. It corresponds to the amplitude used for the excitation transfer and remote entanglement presented on Fig.~3 and Fig.~4. We chose to present data at different pump amplitudes on Fig.~2 as it was recorded with more detail and on a larger range. \\

On Fig.~S7a-b, we plot the recorded Rabi oscillation for Alice and Bob at the maximum drive strength $\xi_2$ that is used in the experiment. On Fig.~S7c-d, we represent the extracted Rabi frequency when varying $\xi_2$. Plain lines are linear fits whose slopes are used as calibration for the transfer and remote entanglement experiments. These slopes are here again in quantitative agreement with the predictions from spectroscopy measurement  represented as dashed black lines.

\begin{figure}[]
\includegraphics[scale=0.6]{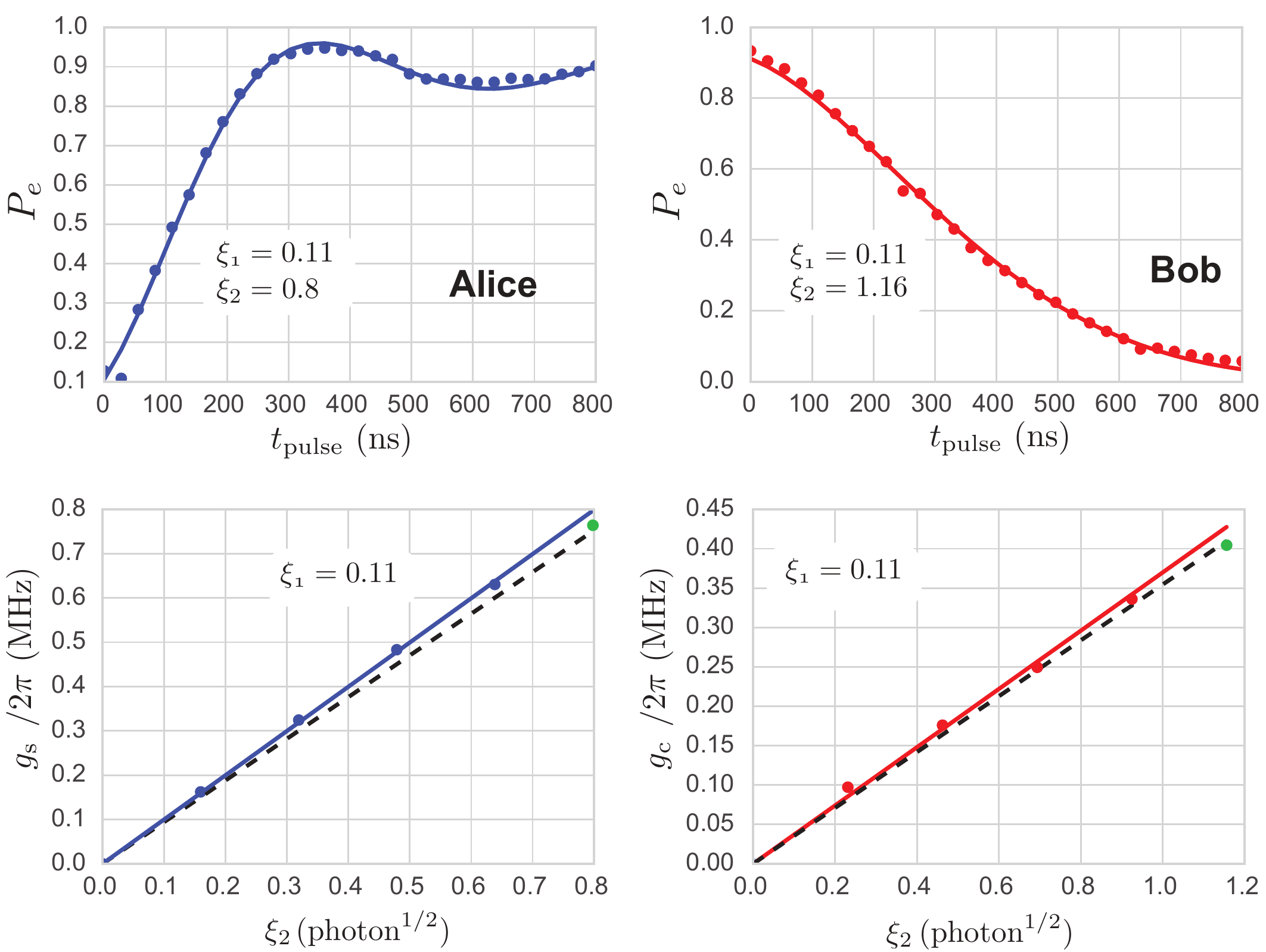}
\caption{\label{fig:rabsup}  \textbf{Top panels} Rabi oscillations when driving a two-photon transition for a varying duration $t_{\mathrm{pulse}}$  are recorded in the qubit excited state populations (dots). Alice is initialized in $|g\rangle$, Bob in $|e\rangle$. The pump amplitude values $\xi_1$ and $\xi_2$ are calibrated through Stark-shift measurements (see section \emph{Stark shift measurement}). Lines  are fits for the two-photon drive strengths $g_\mathrm{s}$ and  $g_\mathrm{c}$. \textbf{Bottom panels} The extracted drive strengths are plotted when varying $\xi_2$ (dots, the green ones are from the top panel fits). Lines are linear fits and their slopes are used as a calibration for release and capture of a shaped photon. Dashed black lines are the drive strengths $|g_{\mathrm{c},\mathrm{s}}|=\chi |\xi_1 \xi_2|$ predicted from some Stark-shift calibration.
}
\end{figure}

\section{Simulation and control pulses computation }

\subsection{Langevin equation}

In this section, we show how the evolution of Alice or Bob can be modeled by coupled Langevin equations. This is useful to obtain an analytical expression for the two-photon Rabi oscillations and for the algorithm computing the control pulses to be applied for pitching or catching  a  shaped wavepacket. For simplicity, we now consider the system made of Bob and its cavity. Alice's case can be directly adapted by relabeling the qubit states as $|g\rangle \leftrightarrow |e\rangle$ and the qubit field operator as $q \leftrightarrow q^{\dagger}$.\\  

Bob's cavity relaxes at rate $\kappa$ and we neglect the qubit relaxation or decoherence, as well as thermal excitation of the two modes. Thus, when starting in $|g0\rangle$ or $|e0\rangle$, the evolution of the system is confined to the three-level manifold spanned by $\{|g0\rangle, |e0\rangle, |g1\rangle\}$  (see Fig.~1). We also assume that the modified resonance condition $\tilde{\omega}_q -\tilde{\omega}_c=\omega_1-\omega_2 - \delta$ is always met. Here we allow a small detuning $\delta$ from the original resonance condition (3b) as this will be useful for capturing or releasing a photon at a frequency slightly different from the one of the cavity. Then, placing ourselves in a frame rotating at $\tilde{\omega}_q$ for the qubit mode and $\tilde{\omega}_c-\delta$ for the cavity mode (the frame where the emitted wavepacket is a real gaussian), the Hamiltonian (1) of the main text boils down to 
\begin{equation}
\frac{H}{\hbar}=\delta c^{\dagger}c+g(t) q^{\dagger}c+g^{\ast}(t)qc^{\dagger}.
\label{hamilsimple}
\end{equation}
Considering that $c$ is damped at a rate $\kappa/2$, we can then write quantum Langevin equations for the operators $q$ and $c$~\cite{gardiner2004quantum} :
\begin{equation}
\begin{split}
\dot{c}&=-i\delta c-ig^{\ast}(t)q-\frac{\kappa}{2}c - \sqrt{\kappa}c_{in}\\
\dot{q}&=-ig(t) c 
\label{langevin}
\end{split}
\end{equation}
where $c_{in}$ is the incoming field at the cavity port (we consider here only the strongly coupled port) linked to the outgoing field by the input-output relation
\begin{equation}
\sqrt{\kappa} c = c_{out}-c_{in}
\end{equation}
These equations are linear and if the system starts in a product of coherent states and the drive is classical, it stays in a product of coherent states. Note that our model would however break down if we were to consider states with a non negligible average number of photon as we have omitted the non-linear part of the Hamiltonian.\\

We now derive an analytical expression for the excited state population of the transmon in presence of a constant two-photon drive and no input field. This is the case for the measurement presented on Fig.~2 of the main text and on Fig.~S7. For simplicity, we take $g$ to be real. From Eq.~\eqref{langevin}, we can show that if $q$ is in the  coherent state $|\alpha_0\rangle$ and  $c$ in the vacuum at $t=0$, at a given time $t$, they are in the coherent state $|\alpha(t)\rangle|\beta(t)\rangle$ with 
\begin{equation}
\begin{split}
\alpha(t)&=\alpha_0~\mathrm{e}^{-\frac{\gamma t}{4}}\left(\mathrm{cosh}\frac{\lambda t}{4}~+~\frac{\gamma}{\lambda} \mathrm{sinh}\frac{\lambda t}{4}\right)~\overset{\mathrm{def}}{=}\alpha_0 f(t)\\
\beta(t)&=\alpha_0\frac{-4 i g }{\lambda}\mathrm{e}^{-\frac{\gamma^{\ast}t}{4}} \mathrm{sinh}\frac{\lambda t}{4}
\label{coherents}
\end{split}
\end{equation}
where $\gamma=\kappa+2i\delta$ and  $\lambda=\sqrt{\gamma^2-16 g^2}$.\\

 Let us now consider the initial state $|\psi_0\rangle = \sqrt{1-\epsilon}|g0\rangle + \sqrt{\epsilon}|e0\rangle$. $|g0\rangle$ is a stationary point of the system dynamics and as mentioned above, the system remains in a three-level manifold so that its state at a later time can be written as $|\psi(t)\rangle = \sqrt{1-\epsilon}|g0\rangle + \sqrt{\epsilon}(\sqrt{\mu(t)}|e0\rangle+\sqrt{1-\mu(t)}|g1\rangle)$. We do not write phases explicitly here as we are only interested in populations. Then the average occupation of the excited state of the transmon is $\epsilon \mu(t)$.\\
If we consider the case $\epsilon \ll 1$, then we can re-write $|\psi_0\rangle=|\alpha_0\rangle_q |0\rangle_c$ where $|\alpha_0\rangle$ denotes a coherent state with $\alpha_0^2=\epsilon$. Then, from Eq.~\eqref{coherents}, the average occupation of the excited state of the transmon at time $t$ is $\alpha_0^2 f(t)^2$. We can then identify $\mu(t)=f(t)^2$. \\
In the case of the experiment where $\epsilon=1$, we then find that 
\begin{equation}
P(|e\rangle)(t)=f(t)^2=\mathrm{e}^{-\frac{\gamma t}{2}}\left(\mathrm{cosh}\frac{\lambda t}{4}~+~\frac{\gamma}{\lambda} \mathrm{sinh}\frac{\lambda t}{4}\right)^2.
\end{equation}
This expression is used to fit the data of Fig.~2 and Fig.~S7. The detected average occupation $P_e$ of the $|e\rangle$ level has a slightly lower contrast due to finite readout fidelity (see Table I).

\subsection{Control pulses generation algorithm}
\label{algo}
We here describe the algorithm  used to compute the control sequences $\{g_s(t)\}_{0<t<T}$ and $\{g_c(t)\}_{0<t<T}$ used for releasing and catching an excitation and for remote entanglement. We here neglect every system imperfections. We first focus on the \emph{catch} sequence by Bob.\\

Following Cirac \emph{et al.}~\cite{Cirac1996}, let us  note that if some control sequence leads to efficient pitch and catch of a ``full'' photon in a given wavepacket when Alice is initially in $|g\rangle$, the same sequence applied when Alice is initially in  $\sqrt{\epsilon}|g\rangle +\sqrt{1-\epsilon}|e\rangle$  will lead to the transfer of a photon in the same wavepacket with amplitude $\sqrt{\epsilon}$, coherently superposed with the system remaining idle in $|e\rangle$ with amplitude $\sqrt{1-\epsilon}$. When $\epsilon \ll 1$, from Bob's point of view, this means catching a small coherent state. Thus, we can compute the catch sequence considering that the field to be caught is coherent, so that the Langevin equations~\eqref{langevin} can be considered as scalar equations. \\
Following~\cite{Korotkov2011}, we then note that a perfect catch implies that no field is reflected off Bob's cavity so that $c_{out}=0$ at all time. Then, given a wavepacket $c_{in}(t)$ with arbitrary shape and normalized as $\int_0^T |c_{in}|^2=n_{phot}\ll 1$~\footnote{It is convenient to think of this wavepacket as containing less than 1 photon even though any normalization would lead to the same result since the equations of motion are formally linear.}, the input-output relation leads to $c(t)=-c_{in}/\sqrt{\kappa}$ and Eq.~\eqref{langevin} becomes  
\begin{subequations}
\begin{align}
\dot{c}_{in}+i\delta c_{in}-\frac{\kappa}{2}c_{in}&=i g^{\ast }q \label{eom1}\\
ig c_{in}=\sqrt{\kappa}\dot{q}\label{eom2}
\end{align}

\end{subequations}
At each discrete time-step $t$, knowing $q(t)$, one then computes $g(t)$ in order to satisfy Eq.~\eqref{eom1}. We then propagate $q$ to $t+\mathrm{d}t$ with Eq.~\eqref{eom2}, and iterate up to the final time $T$. Note however that this equation diverges if we choose $q=0$ as the initial state : a symmetry allows one to choose arbitrarily the phase of $g$ as long as $q$ is initially displaced with the same phase.  We regularize this formal divergence by giving a small real value $q_0\ll \sqrt{n_{phot}}$ to the field at $t=0$ (in practice we choose $n_{phot} /q_0^2=100$).\\

The case of pitching a photon with Alice can be treated similarly. We find Alice's Langevin equation from Bob's by re-labeling the states $|g\rangle \leftrightarrow |e\rangle$ and the field operator as $q \leftrightarrow q^{\dagger}$. One can again consider pitching a coherent state with a shape $c_{out}$ and containing a small number of photons $n_{phot}$. In practice, we will choose the same shape for Alice's pitch and Bob's catch so that $c_{A,out}(t)=c_{B,in}(t+\tau)$. Here $\tau \ll T$ is the (negligible) propagation delay between Alice and Bob. We then compute $g(t)$ by imposing $c=c_{out}/\sqrt{\kappa}$ at all time $t$. \\
The choice of the initial state with respect to $n_{phot}$ is here crucial. If we choose $q^{\dagger}(0)^2=n_{phot}$, the qubit will end up in $|e\rangle$ and the pitch is full (see Fig.~3 of the main text). If alternatively we set $q^{\dagger}(0)^2=n_{phot}/2$, we get a ``half-pitch'' as used for entangling the two qubits. Note that to avoid diverging control pulses when the qubit nears $|e\rangle$ at the end of the full pitch, we reduce the number of photons in the pulse as $n_{phot}=q^{\dagger}(0)^2*0.99$. This is the exact time symmetric situation as to the one mentioned above when catching a wavepacket with a qubit initially in $|g\rangle$.\\

In practice the shapes of the traveling wavepackets were chosen as a trade-off between fast transfer to minimize decoherence of the qubits and limited sideband pump amplitudes to avoid the non-linear region of the two-photon drive (see Fig.~2) and spurious decoherence of the qubits. Exponential shapes of the form $e^{\Gamma t} \theta(\frac{T}{2}-t)+e^{-\Gamma (t-T/2)} \theta(t-\frac{T}{2})$ (with  $\theta$ a Heavyside step function) were tested with similar transfer performances.

\subsection{Cascaded systems simulation}
Following Gardiner and Zoller~\cite{gardiner2004quantum}, we can write a Lindblad master equation for Alice and Bob linked by a directional relaxation channel of transmission $T$. Neglecting (for now) the population and coherence decays of the qubit, the system total density matrix evolution is governed by
\begin{equation}
\begin{split}
\dot{\rho}=&i \Big[\rho, H_A+H_B+i\frac{\sqrt{T\kappa_A\kappa_B}}{2}(c_A^{\dagger}c_B-c_A c_B^{\dagger}) \Big] \\
 &~~~~~+ \sqrt{T}\mathcal{D}[\sqrt{\kappa_A} c_A+\sqrt{\kappa_B} c_B]\rho + (1-\sqrt{T})\mathcal{D}[\sqrt{\kappa_A} c_A]\rho + (1-\sqrt{T})\mathcal{D}[\sqrt{\kappa_B} c_B]\rho,
\label{lindblad}
\end{split}
\end{equation}
where $\mathcal{D}$ is the Lindblad damping superoperator defined by $\mathcal{D}[L]\rho=L\rho L^{\dagger}-\frac{1}{2}(L^{\dagger} L \rho + \rho L^{\dagger} L)$. The Hilbert space is the product of all 4 mode spaces (2 transmons and 2 resonators) and we assume that each mode contains at most one excitation.  We can then simulate the evolution in the 4-qubit manifold spanned by $\{ |g\rangle_A, |e\rangle_A \} \otimes \{ |0\rangle_A, |1\rangle_A \} \otimes \{ |0\rangle_B, |1\rangle_B \} \otimes \{ |g\rangle_B, |e\rangle_B \}$. \\
The Hamiltonians $H_A$ and $H_B$ then take the form of Eq.~\eqref{hamilsimple}. To take into account the small frequency mismatch $(\tilde{\omega}_{cA}-\chi_A-\tilde{\omega}_{cB})/2\pi=600~\mathrm{kHz}$ we set $\delta_B=0$ and $\delta_A/2\pi=600~\mathrm{kHz}$. This means that we perform the simulations in the frame rotating at $\tilde{\omega}_{cB}$, which is also the frame in which the traveling wavepacket is a real gaussian.\\

This allows us to test the control sequences $\{g_s(t)\}_{0<t<T}$ and $\{g_c(t)\}_{0<t<T}$ returned by the algorithm described in the previous section. For instance, with $T=1$, we find that these sequences lead to populations $P(|e\rangle_A),P(|e\rangle_B)> 98~\%$ at the end of a full pitch. This is consistent with the choice of a wavepacket containing $0.99$~photon in the algorithm to avoid divergences.\\

We can easily include the imperfect channel transmission $T<1$ and some other experimental imperfections such as qubit relaxation and dephasing by including other Lindblad terms on the right side of  Eq.~\eqref{lindblad} as $\Gamma_{1}\mathcal{D}[q]$ and $\Gamma_{\phi}\mathcal{D}[q^{\dagger}q]$. Here, $\Gamma_1=1/T_1$ is the qubit relaxation rate and $\Gamma_{\phi}=1/T_2-1/(2T_1)$ is the pure dephasing rate.\\

Finally, imperfect readout fidelities are included by renormalizing the qubits excited state occupation $P_{sim}$ returned by the simulation  as $P_{norm}= f_e P_{sim} +(1-f_g)(1-P_{sim})$. Here, $f_e$ and $f_g$ are the $|e\rangle$ and $|g\rangle$ readout fidelities given in Table~I. We apply the same normalization when considering the coherences $\langle X\rangle_{meas}$ and $\langle Y \rangle_{meas}$ since experimentally, these are measured by rotating the qubit and then performing a finite fidelity readout.\\

Including all these calibrated experimental imperfections, we reproduce quantitatively all the data presented in the main text and the measured Pauli vector components of the entangled state generated after a ``half'' pitch and catch plotted on Fig.~S8.

\begin{figure}[]
\includegraphics[scale=0.6]{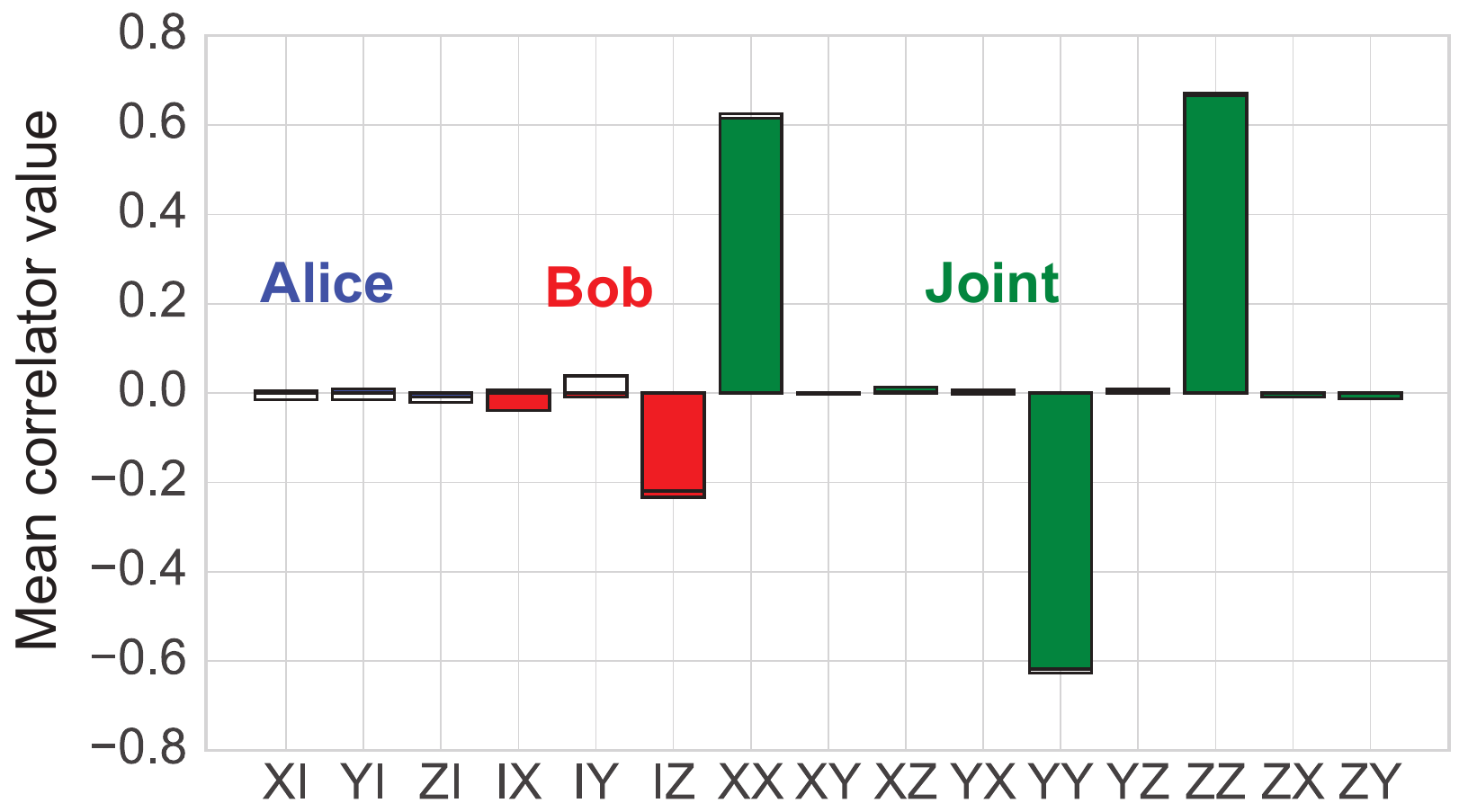}
\caption{\label{fig:correl}  \textbf{Entanglement characterization} Colorbars represent the experimentally measured Pauli vector components of the two-qubit entangled state after a `half' pitch and catch. As on Fig.~4 of the main text, black contours represent the expected values from cascaded quantum system simulations including all calibrated experimental imperfections. The large negative value of $\langle IZ \rangle_{meas}$ mainly originates from photon absorption in the transmission line.
}
\end{figure}

\newpage

\FloatBarrier

\bibliography{bibliography2}{}
\bibliographystyle{unsrt}

\end{document}